\documentclass[12pt]{article}

\usepackage[centertags]{amsmath}
\usepackage{amsmath}
\usepackage{amsfonts}
\usepackage{amssymb}
\usepackage{amsthm}
\usepackage{newlfont}
\usepackage{amscd}
\usepackage{empheq}
\usepackage{epsfig,graphics,graphicx,amsfonts,psfrag}
\usepackage{verbatim}
\usepackage{eucal}

\newcommand{\tr}{\mbox{tr}}
\newcommand{\NN}{{\mathbb N}}

\newcommand{\RR}{{\mathbb R}}
\newcommand{\ZZ}{{\mathbb Z}}

\newcommand{\CC}{{\mathbb C}}
\newcommand{\beq}{\begin{equation}}
\newcommand{\eeq}{\end{equation}}
\newcommand{\ba}{\begin{array}}
\newcommand{\ea}{\end{array}}
\newcommand{\bea}{\begin{eqnarray}}
\newcommand{\eea}{\end{eqnarray}}
\newcommand{\eps}{{\epsilon}}

\usepackage{graphicx}
\usepackage[usenames,dvipsnames]{color}
\usepackage{transparent}
\usepackage{float}

\begin{document}

\begin{center}
  {\bf The effect of loss/gain and hamiltonian perturbations \\ of the Ablowitz - Ladik lattice \\ on the recurrence of periodic anomalous waves}

\vskip 5pt
{\it F. Coppini $^{1,2,3}$ and P. M. Santini $^{1,4}$}

\vskip 5pt

{\it 

$^1$ Dipartimento di Fisica, Universit\`a di Roma "La Sapienza", and 
Istituto Nazionale di Fisica Nucleare (INFN), Sezione di Roma, 
Piazz.le Aldo Moro 2, I-00185 Roma, Italy \\
$^2$ Department of Mathematics, Physics and Electrical Engineering, Northumbria University Newcastle,
Newcastle upon Tyne NE1 8ST, United Kingdom
}
\vskip 5pt

$^{3}$e-mail:  {\tt francesco.coppini@uniroma1.it,\\ francesco.coppini@roma1.infn.it}\\
$^{4}$e-mail:  {\tt paolomaria.santini@uniroma1.it, paolo.santini@roma1.infn.it}
%\bigskip
\vskip 5pt

%{\today}

\end{center}

\begin{abstract}

  The Ablowitz-Ladik (AL) equations are distinguished integrable discretizations of the focusing and defocusing nonlinear Schr\"odinger (NLS) equations. In a previous paper we have studied the effect of the modulation instability of the homogeneous background solution of the AL equations in the periodic setting, showing in particular that both models exhibit instability properties, and studying, in terms of elementary functions, how a generic periodic perturbation of the unstable background evolves into a recurrence of anomalous waves (AWs). Using the finite gap method, in this paper we extend the recently developed perturbation theory for periodic NLS AWs to lattice equations, studying the effect of physically relevant perturbations of the $AL$ equations on the AW recurrence, like: linear loss, gain, and/or Hamiltonian corrections, in the simplest case of one unstable mode. We show that these small perturbations induce $O(1)$ effects on the periodic AW dynamics, generating three distinguished asymptotic patterns. Since dissipation and higher order Hamiltonian corrections can hardly be avoided in natural phenomena involving AWs, and since these perturbations induce $O(1)$ effects on the periodic AW dynamics, we expect that the asymptotic states described analytically in this paper will play a basic role in the theory of periodic AWs in natural phenomena described by discrete systems. The quantitative agreement between the analytic formulas of this paper and numerical experiments is excellent. 
  
\end{abstract}

\section{Introduction}

The Ablowitz-Ladik (AL) equations \cite{AL1,AL2}:
\beq\label{AL}
\ba{l}
i\, \dot{u}_n+u_{n+1}+u_{n-1}-2u_n+\eta |u_{n}|^{2}\left(u_{n-1}+u_{n+1} \right) =0, \ \ \eta=\pm 1, \\
u_n=u(n,t)\in\CC, \ \ \dot{u}_n=\frac{du_n(t)}{dt}, \ \  n\in\ZZ, \ \ t\in\RR,
\ea
\end{equation}
are distinguished examples of integrable nonlinear differential-difference equations reducing, in the natural continuous limit 
\beq\label{limit}
u_n(t)=ih v(\xi,\tau), \ \ h n =\xi, \ \ \tau =h^2 t, \ \ h\to 0, 
\eeq
to the celebrated integrable \cite{ZakharovShabat} nonlinear Schr\"odinger (NLS) equations
\beq\label{NLS}
\ba{l}
iv_{\tau}+v_{\xi\xi}+2\eta |v|^2v=0,  \ \eta=\pm 1, \\
v(\xi,\tau )\in\CC, \ \ v_{\tau}=\frac{\partial v}{\partial \tau}, \ \ v_{\xi\xi}=\frac{\partial^2 v}{\partial {\xi}^2}, \ \ \xi,\tau\in\RR,
\ea
\eeq
where $h$ is the lattice spacing. The two cases $\eta=\pm 1$ distinguish between the focusing ($\eta=1$) and defocusing ($\eta=-1$) NLS regimes. 

The AL equations \eqref{AL} characterize \cite{ItsKorepin} the quantum correlation function  of the XY-model of spins \cite{Lieb}. If $\eta=1$, it is relevant  in the study of anharmonic lattices \cite{Takeno}; it is gauge equivalent to an integrable discretization of the Heisenberg spin chain \cite{ishimori}, and appears in the description of a lossless nonlinear electric lattice ($\eta=1$) \cite{Marquie}. At last, if $\eta=1$, the AL hierarchy describes the integrable motions of a discrete curve on the sphere \cite{Doliwa}.

The well-known Lax pair of equations (\ref{AL}) reads \cite{AL1,AL2}
\beq\label{AL_Lax_1}
\ba{l}
\underline{\psi}_{n+1}(t,\lambda)=L_n(t,\lambda)\underline{\psi}_{n}(t,\lambda), \ \ \ \ {\dot{\underline\psi}_{n}}(t,\lambda)=A_n(t,\lambda)\underline{\psi}_{n}(t,\lambda),\\
L_n(t,\lambda)=\begin{pmatrix}
\lambda & u_n(t) \\[2mm]
-\eta\overline{u}_n(t) & \frac{1}{\lambda}
\end{pmatrix}, \\
A_n(t,\lambda)=i\,\begin{pmatrix}
\lambda^2-1+\eta u_n \overline{u}_{n-1} & \lambda u_n-\frac{u_{n-1}}{\lambda} \\[2mm]
\eta\frac{\overline{u}_n}{\lambda}-\eta\lambda \overline{u}_{n-1}  & 1-\frac{1}{\lambda^2}-\eta {u}_n \overline{u}_{n-1}
\end{pmatrix},
\ea
\eeq
where $\bar f$ is the complex conjugate of $f$, and the matrices $L_n$ and $A_n$ of the Lax pair (\ref{AL_Lax_1}) possess the two symmetry
\beq\label{symmetries}
\ba{l}
L_n(\lambda)=P_\eta\, \overline{L_n\left(\frac{1}{\overline{\lambda}}\right)}P_\eta^\dagger=-\sigma_3\,L_n(-\lambda)\,\sigma_3, \\
A_n(\lambda)=P_\eta\, \overline{A_n\left(\frac{1}{\lambda^*}\right)}P_\eta^\dagger=\sigma_3\,A_n(-\lambda)\,\sigma_3,
\ea
\eeq
where \\
\beq\label{def_P}
\sigma_3=\begin{pmatrix}
  1 & 0 \\
  0 & -1
\end{pmatrix}, \ \ \ \ 
P_\eta=\begin{pmatrix}
0&-\eta\\1&0
\end{pmatrix},
\eeq
implying that, if $\underline{\psi}_n(\lambda,t)=({\psi_1}_n(\lambda,t),{\psi_2}_n(\lambda,t))^T$ is solution of (\ref{AL_Lax_1}), then
\beq\label{symmetries_psi}
\check{\underline\psi}_n(\lambda,t)=\left(
\ba{c}
-\eta\overline{{\psi_2}_n\left(\frac{1}{\overline{\lambda}},t)\right)}\\
\overline{{\psi_1}_n\left(\frac{1}{\overline{\lambda}},t)\right)}
\ea
\right), \ \ \hat{\underline\psi}_n(\lambda,t)=(-1)^n\left(
\ba{c}
{\psi_1}_n(-\lambda,t)\\
-{\psi_2}_n(-\lambda,t)
\ea
\right)
\eeq
are also a solution of (\ref{AL_Lax_1}). The Inverse Scattering Transform of the AL equations \eqref{AL} for localized initial data was developed in \cite{AL1,AL2} (see also \cite{APT}), and for non zero boundary conditions and $\eta=1$ in \cite{Prinari}. The Finite Gap (FG) method \cite{Novikov,Dubrovin,ItsMatveev,Krichever,ItsKotlj} for periodic and quasi-periodic solutions was developed in \cite{Miller}.

Let $\Psi_n(t,\lambda)$ be a fundamental matrix solution of (\ref{AL_Lax_1}) for a periodic  potential $u_n(t)$ of period $N$: $u_{n+N}(t)=u_n(t)$. Then it is well-known that the monodromy matrix
\beq\label{def_monodromy}
T(\lambda,t):=\Psi_{1+N}(t,\lambda)\Psi^{-1}_{1}(t,\lambda)=L_{N}(t,\lambda)L_{N-1}(t,\lambda)\dots L_1(t,\lambda)
\eeq
satisfies the following properties.\\
i) $\det T=\prod\limits_{j=1}^N\left(1+\eta |u_n(t)|^2\right)$ is $t$- and $\lambda$-independent.\\
ii) $\tr~ T$ is a finite Laurent expansion in $\lambda$ with $t$-independent coefficients.\\

Unlike the NLS case, for which the homogeneous background solution is unstable under the perturbation of waves with sufficiently small wave numbers $\kappa$ in the focusing case $\eta=1$ \cite{Talanov,BF,Zakharov,ZakharovOstro}, and always stable in the defocusing case $\eta=-1$, the instability properties of the homogeneous background solution
\beq\label{background}
a \exp(2\,i\,\eta\,|a|^2\,t), \ \ a \ \mbox{complex constant parameter},
\eeq
of the AL equations \eqref{AL} are much richer, and can be summarized as follows.
\beq\label{unstable_cases}
\ba{ll}
\mbox{If } \eta=1, & |\kappa|<\kappa_a:=\arccos\left(\frac{1-|a|^2}{1+|a|^2}\right), \ \forall |a|>0; \\

 \mbox{if } \eta=-1, & |a|>1,\ \forall \kappa,
\ea
\eeq
where $\kappa_a$ is the smallest positive branch of $\arccos$ \cite{Akhm_AL,Otha,CS2}. 

The set of exact AW solutions of the AL equations describing such  nonlinear instability in the case of one and two unstable modes, are presented in \cite{CS2}. In the case of one unstable mode they read
\beq\label{Narita}
{\cal N}(n,t;\kappa,X,T,\rho,\eta)=ae^{2i\eta |a|^2t+i\rho}\, \frac{\cosh\left[\sigma(\kappa) (t-T)+2i\eta\phi\right]+\eta G \cos[\kappa(n-X)]}{\,\cosh[\sigma(\kappa) (t-T)]-\eta G \cos[\kappa(n-X)]}, 
\eeq
where
\beq\label{def_phi}
\cos  \phi= \sqrt{1+\frac{\eta}{a^2} }\,\sin \left(\frac{\kappa}{2}\right), \ \ \ \ 0<\phi<\pi/2,
\eeq
\beq\label{def_G1}
G=\frac{\sin \phi}{\cos\left( \frac{\kappa}{2}\right)},
\eeq
$\kappa$ is the wave number,
\beq\label{def_sigma}
\sigma(\kappa)=2\sqrt{(1+\eta |a|^2)(1-\cos \kappa)\left[(1+\eta |a|^2)\cos \kappa-(1-\eta |a|^2)\right]}
\eeq
is the growth rate of the linearized theory in the unstable cases (\ref{unstable_cases}), 
and $X$, $T$ and $\rho$ are arbitrary real parameters. Since $\phi$ is defined in terms of ($\kappa,a,\eta$) through \eqref{def_phi}, the growth rate $\sigma(\kappa)$ and the parameter $G$ can be expressed in terms of $(\kappa,a,\eta)$ or in terms of $(\phi,a,\eta$) in the following way:
\begin{equation}\label{def_Sigma1}
  \ba{l}
\sigma(\kappa)=2\sqrt{(1+\eta |a|^2)(1-\cos\kappa)\left[(1+\eta |a|^2)\cos\kappa-(1-\eta |a|^2)\right]}\\
=2 |a|^2 \sin( 2\phi ), \\
G=\frac{\sin \theta}{\cos\left( \frac{\kappa}{2}\right)}=\sqrt{1-\frac{\eta}{|a|^2}\frac{1-\cos\kappa}{1+\cos\kappa}}
=\frac{\sqrt{|a|^2+\eta}\sin\phi}{\sqrt{|a|^2\sin^2\phi +\eta}}.
\ea
\end{equation}

If $\eta=1$, (\ref{Narita}) is the Narita solution \cite{Narita} (see also \cite{Akhm_AL}), discrete analogue of the well-known Akhmediev breather (AB) \cite{Akhmed0} solution of focusing NLS. If $\eta=-1$, (\ref{Narita}) is the novel solution describing the MI present if $|a|>1$, $\forall \kappa\in\RR$, unlike the NLS case, and this solution blows up at finite time \cite{CS2}. In this respect we feel that the terminology ``focusing'' and ``defocusing'' NLS equations should not be exported to their AL discretizations, since not only the background solution of the AL equation reducing to the defocusing NLS is unstable under \textit{any} monochromatic perturbation if $|a|>1$, but also such an instability leads generically to blow up at finite time \cite{CS2}. Therefore we prefer to call the AL equations \eqref{AL} for $\eta=\pm 1$ as $AL_{\pm}$ equations, instead of using the terminology ``focusing and defocusing AL equations'', often present in the literature.

The Cauchy problem of the periodic AWs of the focusing NLS equation (\ref{NLS}) has been recently solved in \cite{GS1,GS2}, to leading order and in terms of elementary functions, for generic periodic initial perturbations of the unstable background:
\beq\label{eq:nls_cauchy1}
\ba{l}
v(\xi,0)=a\left(1+\varepsilon w(\xi)\right), \ 0<\varepsilon\ll 1, \ w(\xi+L)=w(\xi), \\
w(\xi)=\sum_{j=1}^{\infty}(c_j e^{i k_j \xi}+c_{-j}e^{-i k_j \xi}),\ \ k_j=\frac{2\pi}{L}j,
\ea
\eeq
in the case of a finite number $N$ of unstable modes, using a suitable adaptation of the finite-gap (FG) method. In the simplest case of a single unstable mode ($N=1$), the above finite gap solution provides the analytic and quantitative description of an ideal Fermi-Pasta-Ulam-Tsingou (FPUT) recurrence \cite{FPU} without thermalization, of periodic NLS AWs over the unstable background (\ref{background}),  described, to leading order, by the AB solution of focusing NLS for the arbitrary real parameters $\tilde a,\rho,k,\tilde X,\tilde T$, but with different parameters at each appearance \cite{GS1}. See also \cite{GS3} for an alternative and effective approach to the study of the AW recurrence in the case of a single unstable mode, based on matched asymptotic expansions; see \cite{GS4} for a finite-gap model describing the numerical instabilities of the AB and \cite{GS5} for the analytic study of the linear, nonlinear, and orbital instabilities of the AB within the NLS dynamics; see \cite{GS6} for the analytic study of the phase resonances in the AW recurrence; see \cite{San} and \cite{CS2} for the analytic study of the FPUT AW recurrence in other NLS type models: respectively the PT-symmetric NLS equation \cite{AM1} and the massive Thirring model \cite{Thirring,Mikhailov}. The AB, describing the nonlinear instability of a single mode, and its $N$-mode generalization were first derived respectively in \cite{Akhmed0}  and \cite{ItsRybinSall}. The NLS recurrence of AWs in the periodic setting has been investigated in several numerical and real experiments, see, e.g., \cite{Yuen1,Yuen3,Kimmoun,Mussot,PieranFZMAGSCDR}, and qualitatively studied in the past via a 3-wave approximation of NLS \cite{Infeld,trillo3}. 

In addition, a perturbation theory describing analytically how the FPUT recurrence of NLS AWs is modified by the presence of a small loss or gain has been recently introduced in \cite{CGS}, showing the these perturbations induce $O(1)$ effects on the periodic AW dynamics, and giving a theoretical explanation of previous interesting real and numerical experiments \cite{Kimmoun,Soto}. The qualitative physical explanation of these analytic results is as follows. In the finite time interval when the AW appears, the non integrable loss/gain perturbation generates a small correction to the unperturbed solution, becoming an $O(1)$ correction when the next AW appears, due to MI. This theory has been applied in \cite{CS1} to the complex Ginzburg-Landau \cite{Newell_Whitehead} and Lugiato-Lefever \cite{LL} models, treated as perturbations of NLS; see also \cite{CGS2}. The case of a Hamiltonian perturbation was postponed to a subsequent paper, since in this case the leading order term in the main formula is identically zero in this case, and one should go to next order in the asymptotic expansion.

These NLS results suggest two interesting problems.\\
i) The construction of the analytic and quantitative description of the dynamics of periodic AWs on lattice models, using as prototype model the AL equations.\\
ii) The understanding of the effects of physically relevant perturbations, like loss, gain, and/or Hamiltonian corrections, on the AW dynamics on lattices.

In \cite{CS2} we concentrated on the first problem. In particular, the periodic Cauchy problem of the AL equations \eqref{AL}, corresponding to a small initial periodic perturbation of the unstable background \eqref{background}
\beq\label{Cauchy}
\ba{l}
u_{n+N}(t)=u_n(t), \ \ \forall n\in\ZZ, \ \forall t\ge 0, \\
u_n(0)=a\left[1+\eps \left(\sum\limits_{j=1}^{p}\left(c_j e^{i k_j n}+c_{-j}e^{-i k_j n}\right)+c_0\right)\right], \ \ \ 0<\eps\ll 1, 
\ea
\eeq
where
\beq
k_j=\frac{2\pi}{N}j, \ \ \ 1\le j \le p, \ \ \ p:=\left\{
\ba{ll}
\frac{N}{2}, & \mbox{if $N$ is even}, \\
\ & \ \\
\frac{N-1}{2}, & \mbox{if $N$ is odd},
\ea\right.
\eeq
has been solved, to leading order and in terms of elementary functions, in the case of one and two unstable modes for the $AL_+$ equation, using the matched asymptotics technique introduced in \cite{GS3}.

If $\eta=1$, the instability condition $|\kappa |<\kappa_a$ \eqref{unstable_cases} implies that the first $M\le p$ modes $\pm\kappa_j,~1\le j \le M$ are unstable, where
\beq\label{M_definition}
M:=\left \lfloor{\frac{N\kappa_a}{2\pi}}\right \rfloor
\eeq
and $\left \lfloor{x}\right \rfloor$ is the largest integer less or equal to $x\in\RR$. If $M=1$ ($2\pi/\kappa_a <N<4\pi/\kappa_a$), the solution of the Cauchy problem \eqref{Cauchy} for $AL_+$ is described, to leading order, by the following expression, uniform in space-time, with $t\le t^{(n)}+O(1)$:
\beq\label{FPUT_AL}
u(x,t)=\sum\limits_{j=0}^n{\cal N}(x,t;\kappa_1,x^{(j)},t^{(j)},\rho_j,1)-ae^{2i|a|^2t}\frac{1-e^{4i\phi_1 n}}{1-e^{4i\phi_1}}+O(\eps).
\eeq
where
\beq\label{def_recurrence}
\ba{l}
x^{(n)}=x^{(1)}+(n-1)\Delta x, \ x^{(1)}=\frac{\arg(\alpha_1)+\pi/2}{\kappa_1}, \ \mod N, \\
t^{(n)}=t^{(1)}+(n-1)\Delta t, \ \ t^{(1)}=\frac{1}{\sigma_1}\log\left(\frac{\sigma^2_1}{2 |a|^2 \eps |\alpha_1|\cos(\kappa_1/2)}\right),\\
\rho_n=2\phi_1+(n-1)4\phi_1,
\ea
\eeq
\beq\label{def_Delta x_Delta t}
\ba{l}
\Delta x=\frac{\arg(\alpha_1\beta_1)}{\kappa_1}, \mod N, \ \ \ \ \ \   
\Delta t=\frac{1}{\sigma_1}\log\left(\frac{\sigma^4_1}{4 |a|^8 \eps^2 |\alpha_1\beta_1|\cos^2(\kappa_1/2)}\right),
\ea
\eeq
and
\beq\label{def_alpha_beta}
\alpha_1:=\overline{c_1}e^{-i\phi_1}-c_{-1}e^{i\eta\phi_1}, \ \ \beta_1=\overline{c_{-1}}e^{i\phi_1}-c_1e^{-i\phi_1}.
\eeq

\begin{figure}[h!!!!]
\centering
\includegraphics[trim=3cm 1cm 0cm 0 ,width=15cm,height=8cm]{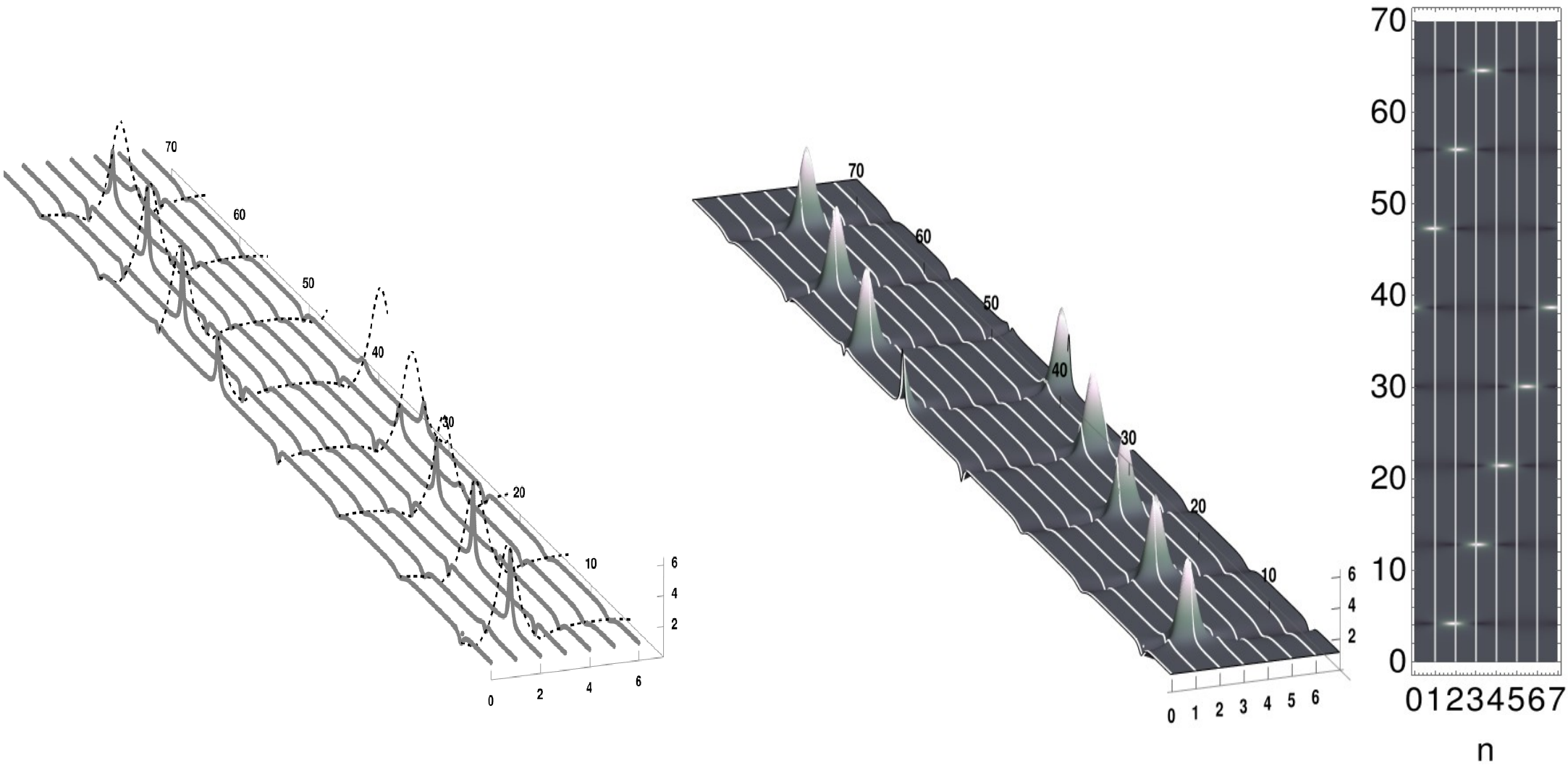}
\caption{3D plot of $|u_n(t)|$ coming from the numerical integration of the Cauchy problem of the AWs for the AL equation ($\eta=1$), in the case of a single unstable mode, using the 6th order Runge-Kutta method \cite{RungeKutta}; the short and long axes are respectively the lattice and time axes. Left: using dotted lines, we add to the figure the Narita solution $|{\cal N}|$ at the max of $|u_n(t)|$, coming from the theory \eqref{FPUT_AL}-\eqref{def_alpha_beta}. Centre: together with the numerical integration (white stripes), we plot the succession of Narita solitons as a continuous plot. Right, the density plot of the central graph. The initial condition $u_n(0)=a(1+\epsilon(c_+e^{ikn}+c_-e^{-ikn}))$, where $N=7$, $a=1.1$, $\epsilon=10^{-4}$, $c_+=0.53-i~0.86$, and $c_-=-0.26+i~0.22$. The sites are defined by lines  parallel to the $t$-axis. }\label{fig:recurrence} 
\end{figure}

This is the analytic and quantitative description of the ideal Fermi-Pasta-Ulam-Tsingou (FPUT) recurrence of AWs of the $AL_+$ equation in term of the initial data through elementary functions. $x^{(1)}$ and $t^{(1)}$ are respectively the first appearance time and the position of the maximum of the absolute value of the AW; $\Delta x$ is the $x$-shift of the position of the maximum between two consecutive appearances, and  $\Delta t$ is the time interval between two consecutive appearances (see Figure~\ref{fig:recurrence}). The quantitative agreement between formulas \eqref{FPUT_AL}-\eqref{def_alpha_beta} and numerical experiments is excellent, being even better than the expected theoretical estimate of one or more orders of magnitude \cite{CS2}!
\iffalse
\begin{document}
	\begin{figure*}[h!!!!]
		\centering
		\includegraphics[trim=0cm 0cm 0cm 0 ,width=13cm,height=8cm]{img/Ablowitz-Ladik.pdf}
		\caption{Left: the 3D plot of $|u_n(t)|$ coming from the numerical integration of the Cauchy problem of the AWs for the AL equation ($\eta=1$), in the case of a single unstable mode, using the 6th order Runge-Kutta method \cite{RungeKutta}. We added to the figure the Narita solutions $|{\cal N}|$ at the max of $|u_n(t)|$  coming from the theory \eqref{FPUT_AL}-\eqref{def_alpha_beta} using dotted lines. Right: the associated density plot. The initial condition $u_n(0)=a(1+\epsilon(c_+e^{ikn}+c_-e^{-ikn}))$, where $M=7$, $a=1.1$, $\epsilon=10^{-4}$, $c_+=0.53-i~0.86$ and $c_-=-0.26+i~0.22$. The sites are defined by lines  parallel to the $t$-axis.} 
              \end{figure*}
              \fi

The same Cauchy problem for the $AL_-$ equation, in the case of one unstable mode, allows one to show analytically how a generic smooth initial perturbation of the background leads to a blow up at finite time described by the solution \eqref{Narita} with $\eta=-1$, whose free parameters are expressed in terms of the initial data via elementary functions \cite{CS2}.

We remark, from (\ref{def_recurrence}), that the maximum of the AW at the first appearance is located on a lattice point $x^{(1)}\in\ZZ$, if the initial data are such that $\frac{\arg\alpha_1}{\kappa_1}+\frac{N}{4}\in\ZZ$; if, in addition, $\Delta x\in\ZZ$, i.e., $\frac{\arg\beta_1}{\kappa_1}-\frac{N}{4}\in\ZZ$, the maxima of the FPUT recurrence are all located in the lattice points.

As in the NLS case, a very distinguished situation occurs when the initial data in \eqref{Cauchy} are such that $|c_1|=|c_{-1}|=:|c|$, corresponding to the case $\alpha_1\beta_1\in\RR$:
\beq
\ba{l}
\alpha_1\beta_1=2 |c|^2\left[\cos\gamma -\cos(2\phi_1) \right]\in\RR, \ \  \ \ \gamma:=\arg(c_1)+\arg(c_{-1}).
\ea
\eeq
Indeed, if $\alpha_1\beta_1>0$, then $\Delta x=0$ and the FPUT recurrence is periodic with period $\Delta t$. If $\alpha_1\beta_1<0$, then $\Delta x=N/2$ and the FPUT recurrence is periodic with period $2 \Delta t$. Therefore, in terms of the initial data:
\beq\label{special}
\ba{l}
|c_1|=|c_{-1}|, \ |\gamma|<2\phi_1 \ \ \ \Leftrightarrow \ \ \ \alpha_1\beta_1>0, \\
|c_1|=|c_{-1}|, \ |\gamma|>2\phi_1 \ \ \ \Leftrightarrow \ \ \ \alpha_1\beta_1<0.
\ea
\eeq
Particularly interesting subcases are $u_n(0)\in\RR$ and $u_n(0)\in i\RR$ \cite{CS2}. \\
a) If $u_n(0)\in\RR$, then $|c_1|=|c_{-1}|, \ \gamma=0$, implying $\alpha_1\beta_1>0$, $\Delta x=0$, and a periodic FPUT recurrence with period $\Delta t$.\\
b) If $u_n(0)\in i\RR$, then $|c_1|=|c_{-1}|, \ |\gamma|=\pi$, implying $\alpha_1\beta_1<0$, $\Delta x=N/2$, and a periodic FPUT recurrence with period $2 \Delta t$.

\textit{Although the conditions \eqref{special} are not generic with respect to the AL dynamics, as we shall see in this paper, they describe the generic asymptotic state when the AL dynamics is perturbed by a small loss or gain, as in the NLS case}. \textit{Since dissipation can hardly be avoided in all natural phenomena involving AWs, and since a small dissipation induces $O(1)$ effects on the periodic AW dynamics, generating the asymptotic states analytically described in this paper, we expect that these asymptotic states, together with their generalizations corresponding to more unstable modes, will play a basic role in the theory of periodic AWs in natural phenomena described by discrete systems}.

In this paper we extend the NLS perturbation theory to lattices the $AL_+$ lattice, in order to show the order one effects of a small perturbation of the equation on the AW dynamics, and exemplify the theory on three basic examples: a linear loss, a linear gain, and a quintic Hamiltonian perturbation.

The paper is organized as follows. In \S 2 we construct the system of gaps corresponding to the background solution \eqref{background}  of the AL equations, and to a generic periodic perturbation of the background solution; in \S 3 we construct the AW perturbation theory in the simplest case of one unstable mode; in \S 4 we apply theis theory to three relevant examples: a linear loss, a linear gain, and a quintic hamiltonian perturbation, enabling one to study quantitatively the order one effects of these perturbations on the AL AW dynamics described by formulas \eqref{FPUT_AL}-\eqref{def_alpha_beta}. The interesting problem of studying the AW dynamics of the physically relevant discrete NLS equations
\beq\label{discreteNLS}
\ba{l}
i\, \dot{u}_n+u_{n+1}+u_{n-1}-2u_n+\eta |u_{n}|^{2}u_{n}=0, \ \ \eta=\pm 1,
\ea
\end{equation}
viewed as a Hamiltonian perturbation of the AL lattices, is postponed to a subsequent paper.

\section{The main spectrum of the perturbed background}

To define the main spectrum associated with the periodic AW Cauchy problem, we find it convenient to work in the gauge defined by
\beq\label{miller_gauge}
\underline{w}_{n}(\lambda,t)=\frac{1}{\prod\limits_{j=0}^{n-1}\sqrt{1+\eta |u_j|^2}}\underline{\psi}_{n}(\lambda,t),
\eeq
corresponding to the Lax pair (see the Appendix of \cite{Miller})
\beq\label{miller_Lax}
\ba{l}
\underline{w}_{n+1}(\lambda,t)=\tilde L_n(t)\underline{w}_{n}(\lambda,t), \ \ \dot{\underline{w}}_{n}(\lambda,t)=\tilde A_n(\lambda,t)\underline{w}_{n}(\lambda,t),\\
\tilde L_n=\frac{1}{\sqrt{1+\eta |u_n|^2}}L_n, \ \ \ \Rightarrow \ \ \det\tilde L_n =1, \\
\tilde A_n=i\begin{pmatrix}\frac{1}{2}\left(\lambda-\frac{1}{\lambda}\right)^2+\frac{\eta}{2}\left(u_n\overline{u}_{n-1}+\overline{u_n}u_{n-1}\right) & \lambda u_n-\frac{u_{n-1}}{\lambda}\\[2mm]-\lambda\eta\, \overline{u}_{n-1}+\eta\frac{\overline{u_n}}{\lambda}&-\frac{1}{2}\left(\lambda-\frac{1}{\lambda}\right)^2-\frac{\eta}{2}\left(u_n\overline{u}_{n-1}+\overline{u_n}u_{n-1}\right)\end{pmatrix}.
\ea
\eeq
The corresponding monodromy matrix $\tilde T$ is simply related to $T$:
\beq
\ba{l}
\tilde T(\lambda,t):=\tilde L_{N}(\lambda,t)\tilde L_{N-1}(\lambda,t)\dots \tilde L_1(\lambda,t)=\frac{1}{\sqrt{\det T}}T(\lambda,t); 
\ea
\eeq
then
\beq\label{tilde_T vs_T}
\det \tilde T=1, \ \ \ \ \ \tr \tilde T=\frac{\tr T}{\sqrt{\det T}}
\eeq
are also $t$-independent.

\iffalse

Let $W(n,\lambda,t_0)$ be any fundamental matrix solution of the first equation in (\ref{miller_lax}); then \beq
\tilde W(n,n_0,\lambda,t_0)=W(n,\lambda,t_0)W^{-1}(n_0,\lambda,t_0)
\eeq
is the fundamental solution satisfying the boundary condition $\tilde W(n_0,n_0,\lambda,t_0)=I$, where $I$ is the identity matrix, and $\tilde W(n+M,n_0,\lambda,t_0)$ is the monodromy matrix. Changing $(n_0,t_0)$, the new monodromy matrix is a similarity transformation of the previous one; therefore its invariants (trace, determinant, etc\dots) do not change.

We also remark that, if $n_0=0$, the monodromy matrix reads
\beq
T(\lambda,t_0)=W(M,\lambda,t_0)W^{-1}(0,\lambda,t_0)=\tilde L_{M-1}(\lambda,t_0)\tilde L_{M-2}(\lambda,t_0)\dots \tilde L_{0}(\lambda,t_0);
\eeq
it is an entire function in $\CC-\{0\}$, and all its invariants are constants of motion with respect to the AL dynamics, with $\det T=1$.

\fi

The  eigenvalues of $\tilde T$:
\beq\label{eigenv_T}
\chi_{\pm}=\frac{\tr \tilde T(\lambda)}{2}\pm\sqrt{\left(\frac{\tr \tilde T(\lambda)}{2}\right)^2 -1}
\eeq
and its eigenvectors live on a two-sheeted covering $\Gamma$ of the $\lambda$ plane, due to the square root in (\ref{eigenv_T}), and play a crucial role in the construction of the Bloch eigenvectors, defined by the equations
\beq
\ba{l}
\vec w^{B,\pm}_{n+1}(\gamma)=\tilde L_n(t)\vec w^{B,\pm}_{n}(\gamma), \\
\ \\
\vec w^{B,\pm}_{n+N}(\gamma)=\chi_{\pm}(\gamma)\vec w^{B,\pm}_{n}(\gamma), \ \ \gamma\in\Gamma ,
\ea
\eeq
where the Floquet multipliers $\chi_{\pm}(\gamma)$ are just the eigenvalues of the monodromy matrix $\tilde T$. The main spectrum is defined by the condition
\beq\label{def_main_spectrum}
|\chi(\gamma)|=1 \ \ \ \ \Leftrightarrow \ \ \ \ -2\le \tr \tilde T(\lambda)\le 2,
\eeq
and, generically, consists of disconnected curves (the bands) in the complex $\lambda$ plane. The end points of the main spectrum, corresponding to periodic and anti-periodic Bloch eigenvectors and characterized by the conditions
\beq
\chi(\gamma)=\pm 1 \ \ \ \ \Leftrightarrow \ \ \ \ \tr \tilde T=\pm 2,
\eeq
are the branch points and the multiple points (arising from the coalescence of two or more branch points). These end points are gauge independent and, in the original gauge (\ref{AL_Lax_1}), they are characterized instead by the condition
\beq\label{spectrum_gauge_AL}
\tr T=\pm 2 \sqrt{\det T}.
\eeq

We remark that the symmetries (\ref{symmetries}) imply the relations
\beq
\tr \tilde T(\lambda)=(-\eta)^N \tr \tilde T(1/\overline{\lambda})=-\tr \tilde T(-\lambda).
\eeq
Consequently the main spectrum and its end points are invariant under the transformations $\lambda\to 1/\overline{\lambda}$ and $\lambda\to -\lambda$ (if $\lambda$ belongs to the main spectrum, also $1/\overline{\lambda},-\lambda,-1/\overline{\lambda}$ belong to the main spectrum). In particular, if $\lambda$ is a branch point or a multiple point, then also $1/\overline{\lambda},-\lambda,-1/\overline{\lambda}$ are rispectively branch points or multiple points.

\subsection{The main spectrum of  the background}.

For the background solution $u^{[0]}_n(t)=a e^{2i\eta |a|^2t}$ the above quantities are explicit. The n-periodic matrix fundamental solution of the Lax pair (\ref{miller_Lax}) reads:
\begin{equation}\label{sol_miller_gauge} 
W^{[0]}_n(t)=e^{i\left(\eta |a|^2t+\frac{\arg a}{2}\right)\sigma_3}
\begin{pmatrix}e^{s(\lambda)\,\exp(i\eta\,\mu)\,t}e^{i\,\eta\,\mu\, n}&&e^{s(\lambda)\,\exp(-i\eta\mu)\,t}e^{-i\,\eta\,\mu\, n}\\g_+(\lambda)\,e^{s(\lambda)\,\exp(i\,\eta\,\mu)\,t}\,e^{i\,\eta\,\mu\, n}&&
g_-(\lambda)\,e^{s(\lambda)\,\exp(-i\,\eta\,\mu)\,t}e^{-i\,\eta\,\mu\, n}
\end{pmatrix},
\end{equation}
where $\lambda$ and $\mu$ satisfy the equation
\beq\label{RS}
\cos\mu=\frac{1}{2\sqrt{1+\eta |a|^2}}\left(\lambda+\lambda^{-1} \right),
\eeq
and
\beq\label{def_g_s}
\ba{l}
g_{\pm}(\lambda)=\frac{1}{|a|}\left(-\frac{1}{2}\left(\lambda-\frac{1}{\lambda}\right)\pm\sqrt{\frac{1}{4}\left(\lambda-\frac{1}{\lambda}\right)^2 -\eta |a|^2} \right), \\
s(\lambda)=i\sqrt{1+\eta|a|^2}\left(\lambda-\frac{1}{\lambda}\right).
\ea
\eeq
From equations (\ref{RS}) and (\ref{def_phi}) it follows that
\beq\label{aa}
\ba{l}
\sin\phi=\frac{1}{2 |a|\sqrt{\eta}}\left(\lambda-\frac{1}{\lambda}\right), \ \ \ 
\cos\phi=\frac{1}{|a|}\sqrt{|a|^2-\eta\left(\frac{\lambda-\lambda^{-1}}{2}\right)^2}, \\
\sin\mu =\frac{\sqrt{\eta |a|^2-\left(\frac{\lambda-\lambda^{-1}}{2}\right)^2}}{\sqrt{1+\eta |a|^2}},
\ea
\eeq
implying 
\beq
g_{\pm}(\lambda)=\pm \frac{i}{\sqrt{\eta}}\, e^{\pm i\phi},
\eeq
and
\beq\label{Bloch_2}
\ba{l}
W^{[0]}_n(t)=e^{i\left(\eta |a|^2t+\frac{\arg a}{2}\right)\sigma_3}
\begin{pmatrix}e^{i\,\eta\,\mu\, n}&&e^{-i\,\eta\,\mu\, n}\\
\frac{i}{\sqrt{\eta}}e^{i\,\eta\,\left(\mu\, n +\phi\right)}   &&
-\frac{i}{\sqrt{\eta}}e^{-i\,\eta\,\left(\mu\, n +\phi\right)}
\end{pmatrix} e^{-\frac{\sigma}{2}t \sigma_3+i\nu\,t}, \\
\nu=2|a|\sqrt{\eta+|a|^2}\sin\phi \cos\mu\,t .
\ea
\eeq
Then the monodromy matrix reads
\beq
\ba{l}
\tilde T^{[0]}(\lambda,t)=W^{[0]}_{N+1}(t)\left(W^{[0]}_1\right)^{-1}(t)=W^{[0]}_{N}(t)\left(W^{[0]}_0\right)^{-1}(t) \\
=\frac{1}{\cos\phi}\begin{pmatrix}
\cos\left(\mu N-\phi \right) & \frac{1}{\sqrt{\eta}}\sin\left(\mu N\right)e^{i\left(2\eta |a|^2 t+\arg a \right)} \\
-\sqrt{\eta}\sin\left(\mu N\right)e^{-i\left(2\eta |a|^2 t+\arg a \right)} & \cos\left(\mu N+\phi \right)
\end{pmatrix}
\ea
\eeq
and
\beq
\tr\tilde T^{[0]}(\lambda)=2\cos(\mu N).
\eeq

The definition (\ref{def_main_spectrum}) of the main spectrum implies that $\mu\in\RR$ and, from (\ref{RS}):
\beq\label{main_spectrum}
\ba{ll}
\{\lambda\in\RR, \ 1/\lambda_0 \le |\lambda |\le\lambda_0\}\cup\{\lambda\in\CC, \ |\lambda|=1 \}, & \mbox{if }\eta=1, \\
\{\lambda=e^{i\theta}, \ \theta\in\RR, \ \pi-\theta_m\le|\theta |\le\theta_m  \}, & \mbox{if }\eta=-1, \ |a|<1, \\
\{\lambda=i p, \ p\in\RR, \ 1/p_0\le p\le p_0  \}, & \mbox{if }\eta=-1, \ |a|>1,
\ea
\eeq
where
\beq
\lambda_0=\sqrt{1+|a|^2}+|a|, \ \ \theta_m=\arccos\left(\sqrt{1-|a|^2}\right), \ \ p_0=\sqrt{|a|^2 -1}+|a|.
\eeq

\begin{figure*}[h]
	\centering
	\includegraphics[width=12cm
        % ,height=10cm
        ]{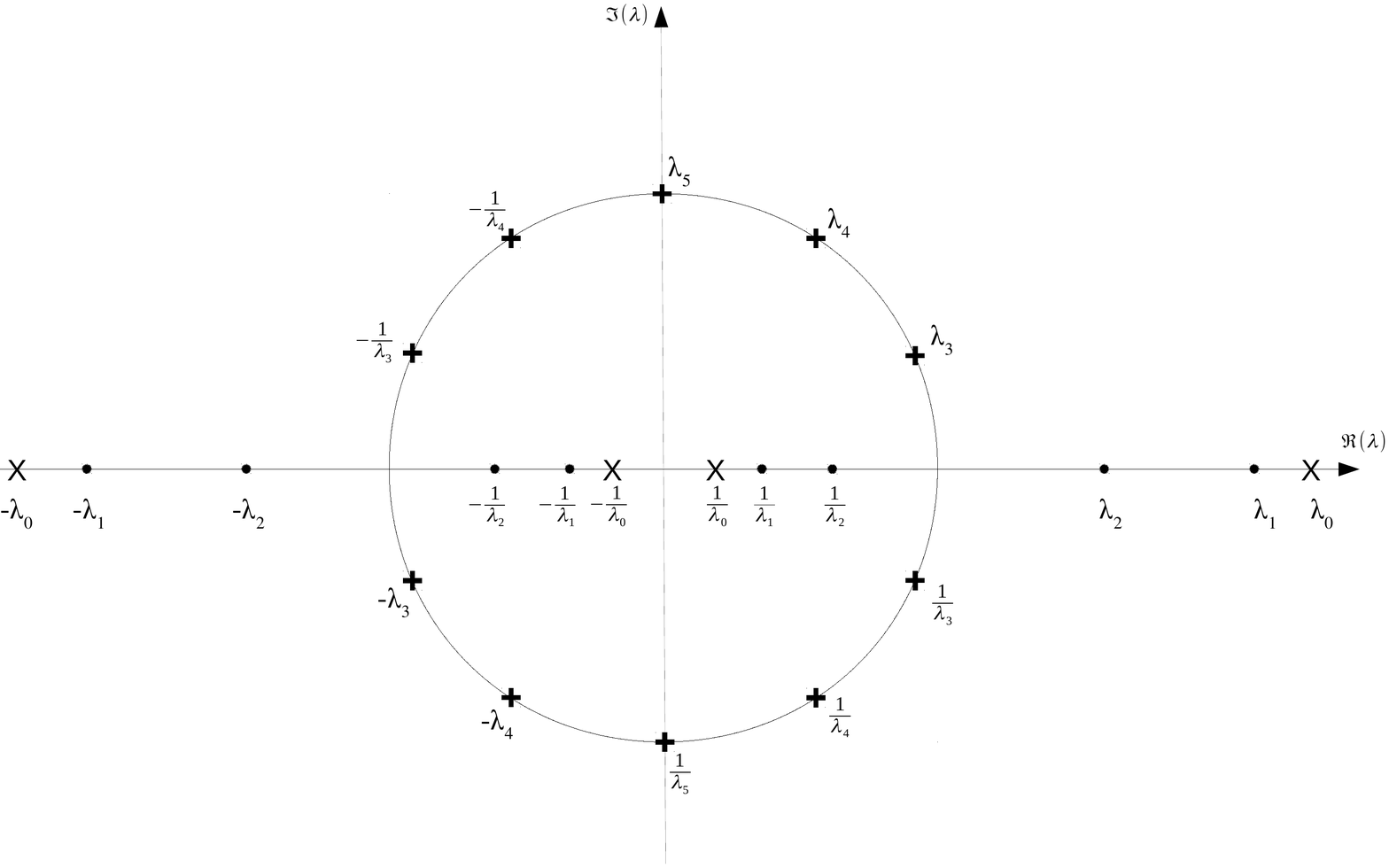}
	\caption{Spectrum in the complex $\lambda$ plane associated with the background solution of the AL equation, with periodicity N=10 and $a=1$. The double points of the spectrum associated with the unstable modes are the black points on the real axis, those associated with the stable modes, indicated by the symbol "+", are located on the unit circle. The four branch points are real and indicated by the symbol "X". }\label{spectrum1}
\end{figure*}

The end points of the main spectrum, corresponding to $\tr\tilde T^{[0]}=\pm 2$, are characterized by
\beq
\mu_n=\frac{\pi}{N}n, \ \ \ 0\le n\le N .
\eeq
From (\ref{RS}), for a given $\mu_n, \ 0\le n\le N$, we have two values $\lambda^{\pm}_n$ of $\lambda$, corresponding to the $2N+2$ end points
\beq\label{def_lambdapm}
\lambda^{\pm}_n=\sqrt{1+\eta |a|^2}\cos\mu_n \pm\sqrt{(1+\eta |a|^2)\cos^2\mu_n -1}, \ \ 0\le n\le N ,
\eeq
with the following symmetries:
\beq\label{symm}
\lambda^{-}_n=1/\lambda^{+}_n =-\lambda^{+}_{N-n}, \ \ 0\le n\le N.
\eeq
Since
\beq
\ba{l}
\partial_{\lambda}\tr\tilde T_0(\lambda)=\frac{N}{\sqrt{1+\eta |a|^2}}\frac{(1-\lambda^{-2})\sin(\mu N)}{\sin\mu}, \\
\partial^2_{\lambda}\tr\tilde T_0(\lambda)=
\frac{N}{\sqrt{1+\eta |a|^2}}\left\{2\frac{\sin(\mu N)}{\lambda^3\sin\mu}+(1-\lambda^{-2})\left[N\frac{\cos(\mu N)}{\sin\mu}-\frac{\sin(\mu N)\cos\mu}{\sin^2(\mu)} \right] \right\},
\ea
\eeq
it follows that
\beq
\partial_{\lambda}\tr\tilde T_0(\lambda)\Big|_{\lambda^{\pm}_0}\ne 0
, \ \ \partial_{\lambda}\tr\tilde T_0(\lambda)\Big|_{\lambda^{\pm}_N}\ne 0, 
\eeq
implying that the four points $\lambda^{\pm}_0$, $\lambda^{\pm}_N$, are branch points (indicated by the symbol ``X'' in Figure~\ref{spectrum1}); it also follows that  
\beq
\ba{l}
\partial_{\lambda}T_0(\lambda)\Big|_{\lambda^{\pm}_n}=0, \\
\partial^2_{\lambda}T_0(\lambda)\Big|_{\lambda^{\pm}_n}=\frac{N^2}{\sqrt{1+\eta |a|^2}}\frac{(-1)^n}{\sin\mu_n}(1-(\lambda^{\pm}_n)^2)\ne 0, \ \ \ n\ne 0,N,
\ea
\eeq
implying that the remaining end points are double points (the black points and those indicated by the symbol ``+'' in Figure~\ref{spectrum1}).

We have the following picture depending on $\eta$ and $|a|$.\\
i) $\eta=1$. The symmetries (\ref{symm}) suggest the following numeration rule
\beq
\lambda_n:=\lambda^{+}_n, \ \ \ 0\le n\le \left\lfloor{\frac{N}{2}}\right\rfloor ,
\eeq
where
\beq
\ba{lll}
\lambda_n>1, & \lambda_n>\lambda_{n+1}, & 0\le n\le M, \\
\lambda_n=e^{i\arg\lambda_n}, & \arg\lambda_n<\arg\lambda_{n+1}, & M+1\le n\le \left\lfloor{\frac{N}{2}}\right\rfloor , 
\ea
\eeq
where  M is the number of unstable modes, as in \eqref{M_definition}.

%\beq
%n_c:=\left\lfloor{\frac{N}{\pi}\arccos\left(\frac{1}{\sqrt{1+|a|^2}} \right)}\right\rfloor .
%\eeq
Then the remaining end points are the sets
\beq
\{1/\lambda_n, \ 0\le n\le \left\lfloor{\frac{N}{2}}\right\rfloor\}, \{-\lambda_n, \ 0\le n\le \left\lfloor{\frac{N}{2}}\right\rfloor\}, \{-1/\lambda_n, \ 0\le n\le \left\lfloor{\frac{N}{2}}\right\rfloor\}.
\eeq
If $N$ is even, $\lambda_{\left\lfloor{\frac{N}{2}}\right\rfloor}=\lambda_{\frac{N}{2}}=i$, and in this case the four points reduce to two, since $-\lambda_{\frac{N}{2}}=1/\lambda_{\frac{N}{2}}$.

We remark that the reality condition $(1+|a|^2)\cos^2\mu_n >1$ for the end points in (\ref{def_lambdapm}) is equivalent to the instability condition (\ref{unstable_cases}), with $\kappa_n=2\mu_n$, Therefore the real double points
\beq
\ba{l}
\{\lambda_n,1/\lambda_n,-\lambda_n,-1/\lambda_n, \ 1\le n\le n_c \}, 
\ea
\eeq
correspond to the unstable modes (the black points in Figure~\ref{spectrum1}), while the remaining double points, located on the unit circle (the points indicated by ``+''), correspond to the stable modes. 

\subsection{The main spectrum of  a periodically perturbed background}

Here we concentrate on the construction of the end points of the main spectrum associated with a periodic perturbation of the background. As we have seen, apart from the branch points $\lambda_0,1/\lambda_0,-\lambda_0,-1/\lambda_0$, the remaining end points of the main spectrum of the background are double points. Therefore a generic periodic perturbation of the background will resolve such a degeneration in agreement with the following basic principles of perturbation theory for doubly degenerate eigenvalues.

Consider the eigenvalue equation for the unperturbed operator ${\cal L}_0$:
\beq
{\cal L}_0 |\psi>=\lambda |\psi>,
\eeq
where $\lambda$ is a doubly degenerate eigenvalue corresponding to the two independent eigenfunctions $|f^{\pm}>$. Then the perturbed operator ${\cal L}={\cal L}_0+\eps {\cal L}_1$ resolves the degeneration in the following way:
\beq
{\cal L}|\psi,\pm>=\left(\lambda+\eps \Delta^{\pm}\right)|\psi,\pm>,
\eeq
where $\Delta^{\pm}$ are the eigenvalues of the $2\times 2$ matrix 
\beq
V=\begin{pmatrix}
<f^{+}|{\cal L}_1 |f^{+}\;> & <f^{+}|{\cal L}_1 |f^{-}> \\
<f^{-}|{\cal L}_1 |f^{+}> & <f^{-}|{\cal L}_1 |f^{-}>
\end{pmatrix},
\eeq
made of the matrix elements of the perturbation ${\cal L}_1$ with respect to the invariant subspace $\{|f^{-}>,|f^{+}>\}$ associated with the eigenvalue $\lambda$, $<f^{\pm}|$ are the corresponding elements of the dual basis such that
\beq
<f^{-}|f^{-}>=<f^{+}|f^{+}>=1, \ \ <f^{\mp}|f^{\pm}>=<f^{\mp}|f>=0,
\eeq
and $|f>$ is any other element of the basis of eigenfunctions of ${\cal L}_0$.

To apply this result to our case, we first observe that the first equation in (\ref{miller_Lax}) (the spectral problem) can be written as the eigenvalue problem \cite{Miller}
\beq
{\cal L}_n\vec w_n = \lambda\vec w_n, \ \ \ \ {\cal L}_n=\begin{pmatrix}\sqrt{1+\eta|u_n|^2} \textbf{E} & -u_n\\
-\eta \overline{u}_{n-1}&\sqrt{1+\eta|u_{n-1}|^2}\textbf{E}^{-1}
\end{pmatrix}, 
\eeq
where \textbf{E} is the forward lattice shift: $\textbf{E}f_n =f_{n+1}$.  Let us specialize the previous formulas to the case $\eta=1$ and compute the effect of the initial monochromatic perturbation
\beq
u_{n}=a\left( 1+\eps(c_j e^{ik_j n}+c_{-j}e^{-i\kappa_j n})\right)
\eeq
on the spectrum discussed in the previous section.

Then ${\cal L}={\cal L}_0+\eps {\cal L}_1 +O(\eps^2)$, where
\beq
{\cal L}_0=\begin{pmatrix}
\sqrt{1+|a|^2}\textbf{E}& -a \\[1mm]
-\bar a &  \sqrt{1+|a|^2}\textbf{E}^{-1}
\end{pmatrix},
\eeq
{\scriptsize\begin{equation*}
{\cal L}_1=\begin{pmatrix}\frac{|a|^2}{2\sqrt{1+|a|^2}} \left[(c_{j}+\overline{c_j})e^{2i\mu_j n}+(c_{-j}+\overline{c_{j}})e^{-2i\mu_j n}\right]\textbf{E}& -a\left(c_{j}e^{2i\mu_j n}+c_{-j}e^{-2i\mu_j n}\right)\\[1mm]
-\overline{a}(\overline{c_{j}}e^{-2i\mu_j(n-1)}+\overline{c_{-j}}e^{2i\mu_j (n-1)}) &\frac{|a|^2}{2\sqrt{1+|a|^2}} \left[(c_{j}+\overline{c_{-j}})e^{2i\mu_j (n-1)}+(c_{-j}+\overline{c_{j}})e^{-2i\mu_j (n-1)}\right]\textbf{E}^{-1}
\end{pmatrix}\;.
\end{equation*}}\\[2mm]
We use the following notation for the Bloch eigenfunctions (\ref{Bloch_2}) at $t=0$, in the generic double points $\lambda_j, \ 1/\lambda_j$:\\
\begin{equation*}
|\lambda_{j},\pm>=e^{i\frac{\arg a}{2}\sigma_3}\begin{pmatrix}1\\[3mm]
\pm i e^{\pm i\phi_j}\end{pmatrix}e^{\pm i\mu_j n}\;;\hspace{1cm}|\lambda^{-1}_j,\pm>=e^{i\frac{\arg a}{2}\sigma_3}\begin{pmatrix}1\\[3mm]
\pm i e^{\mp i\phi_j}\end{pmatrix}e^{\pm i\mu_j n}\;;
\end{equation*}
where 
\beq\label{def_phi_j}
\cos  \phi_j= \frac{\sqrt{1+|a|^2}}{|a|}\,\sin\mu_j, \ \ \ \ \kappa_j=2\mu_j .
\eeq

Then
\begin{equation*}
\ba{l}
V(\lambda_j)=\begin{pmatrix}<+,\lambda_j|\,{\cal L}_1\,|\lambda_j,+>& <+,\lambda_j|\,{\cal L}_1\,|\lambda_j,->\\[3mm]
<-,\lambda_j|\,{\cal L}_1\,|\lambda_j,+>&<-,\lambda_j|\,{\cal L}_1\,|\lambda_j,->\end{pmatrix}\\ =
\frac{|a|\lambda_j}{2\sqrt{1+|a|^2}\sin\phi_j}\begin{pmatrix}0 & \beta_j e^{-i\mu_j-i\phi_j}\\ \alpha_je^{i\mu_j+i\phi_j} & 0\end{pmatrix},
\ea
\end{equation*}
its eigenvalues are $\Delta^{\pm}_j=\pm \frac{|a|}{2\sqrt{1+|a|^2}}\frac{\lambda_j\sqrt{\alpha_j\beta_j}}{\sin\phi_j}$, and the double point $\lambda_j$ splits into the two branch points
\beq
E^{\pm}(\lambda_j)=\lambda_j\left(1\pm\eps \frac{|a|}{2\sqrt{1+|a|^2}}\frac{\sqrt{\alpha_j \beta_j}}{\sin\phi_j}+O(\eps^2)\right),
\eeq
generating the gap
\beq\label{gap}
E^+(\lambda_j)-E^-(\lambda_j)=\frac{\eps |a|\lambda_j \sqrt{\alpha_j\beta_j}}{\sqrt{1+|a|^2}\sin\phi_j}.
\eeq

Proceeding in a similar way, one constructs the matrices $V(\lambda^{-1}_j)$, $V(-\lambda_j)$,$V(-\lambda^{-1}_j)$; for instance
\begin{equation*}
\begin{split}
V\left(\lambda^{-1}_j\right)=-\frac{|a|}{2\sqrt{1+|a|^2}\sin\phi_j\lambda_j}\begin{pmatrix}0&&e^{i\phi_j-i\mu_j}\,\tilde{\beta}_j\\[3mm]
e^{-i\phi_j+i\mu_j}\,\tilde{\alpha}_j && 0 \end{pmatrix},
\end{split}
\end{equation*}
where 
\begin{equation*}
\tilde{\alpha}_j=\overline{c_j}e^{i\phi_j}-c_{-j}e^{-i\phi_j};\hspace{1cm} \tilde{\beta}_j=\overline{c_{-j}}e^{-i\phi_j}-c_j e^{i\phi_j}.
\end{equation*}
Therefore the double point $\lambda^{-1}_j$ splits into the two branch points
\beq
E^{\pm}(\lambda^{-1}_j)=\lambda^{-1}_j\left(1\mp \frac{\eps |a|}{2\sqrt{1+|a|^2}}\frac{\sqrt{\tilde{\alpha}_j \tilde{\beta}_j}}{\sin\phi_j} +O(\eps^2)\right).
\eeq

\begin{figure*}[h!!!!!!!!]
	\centering
	\includegraphics[%trim=0cm 0cm 0cm 0,
        width=11.5cm%,height=10cm
        ] {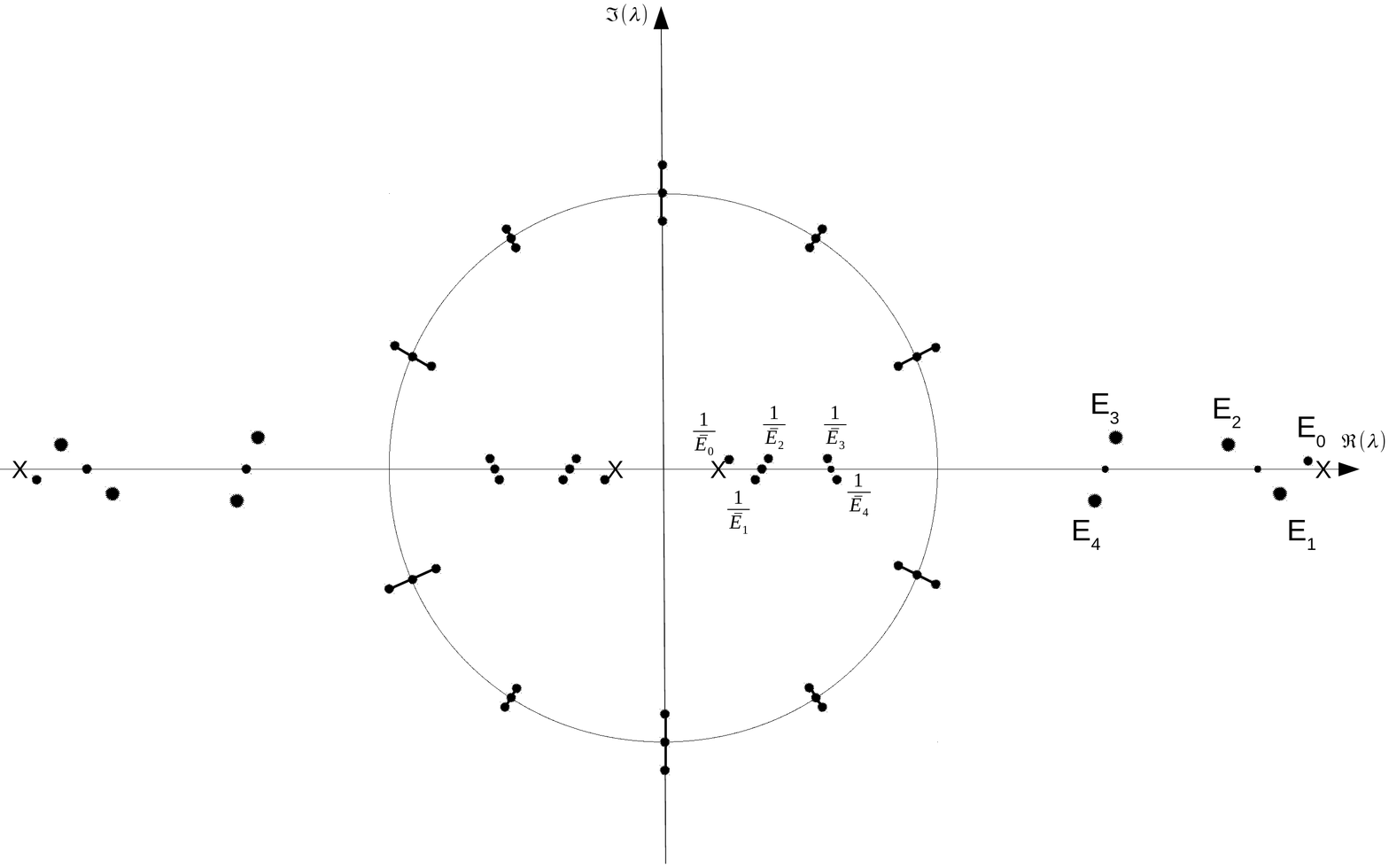}
	\caption{The effect of an initial generic periodic perturbation on the spectrum of  Figure~\ref{spectrum1}. The degenerate double points  are now open gaps. The ones corresponding to the double points on the unit circle are radial; the ones corresponding to double points on the real axis have arbitrary inclination. } \label{spectrum2}
\end{figure*}

If $\lambda_j\in\RR$ (the corresponding mode is unstable), the splitting is generic: $\alpha_j\beta_j$ is an arbitrary complex parameter and the corresponding gap $E^+-E^-$ has an arbitrary inclination and length. In addition, since $\tilde\alpha_j\tilde\beta_j=\overline{\alpha_j\beta_j}$, the branch points associated with  $\lambda_j$ and $\lambda^{-1}_j$ satisfy the symmetry relation
\beq
E^{\pm}(\lambda^{-1}_j)=\frac{1}{\overline{E^{\pm}(\lambda_j)}}.
\eeq
If $\lambda_j$ is on the unit circle: $\lambda_j=e^{i\arg\lambda_j}$ (the corresponding mode is stable), then $\phi_j\in i\RR$ and  
\beq
\ba{l}
\cos\mu_j=\frac{\cos(\arg\lambda_j)}{\sqrt{1+|a|^2}}, \ \ \ \ \sin\phi_j=i\frac{\sin(\arg\lambda_j)}{\sqrt{1+|a|^2}}, \ \ \ \ \beta_j=-\overline{\alpha_j};
\ea
\eeq
therefore
\beq
E^{\pm}(\lambda_j)=\lambda_j \left(1\pm \frac{\eps |a|}{2\sqrt{1+|a|^2}}\frac{|\alpha_j |}{\sin(\phi_j)} +O(\eps^2)\right).
\eeq
It follows that the corresponding branch points are on the line from the origin to $\lambda_j$ (the splitting is radial), and satisfy the relation (see Figure~\ref{spectrum2})
\beq
E^{-}(\lambda_j)=\frac{1}{\overline{E^{+}(\lambda_j)}}.
\eeq

      We finally remark that, since the product $\alpha_1\beta_1$ plays a basic role in the FPUT recurrence of AL AWs, appearing in the definitions (\ref{def_Delta x_Delta t}) of $\Delta x$ and $\Delta t$, and since such a product appears also in the definition (\ref{gap}) of the gap $E^+(\lambda_1)-E^-(\lambda_1)$, it is possible to express $\Delta x$ and $\Delta t$ in terms of the gap, obtaining the spectral representation of the FPUT recurrence:
\beq\label{spectral_recurrence}
\ba{l}
\Delta x=\frac{2}{\kappa_1}\arg\left(E^+(\lambda_1)-E^-(\lambda_1)\right), \\
\Delta t=\frac{1}{\sigma_1}\log\left(\frac{\sigma_1^4 \lambda_1^2}{4|a|^6(1+|a|^2)\sin^2(\phi_1)\cos^2(\kappa_1/2)|E^+(\lambda_1)-E^-(\lambda_1)|^2} \right).
\ea
\eeq

\section{FPUT recurrence of AWs for the perturbed $AL_+$ equation}

The main spectrum is a constant of motion with respect to the $AL_+$ time evolution. If one perturbs such evolution:
\beq
\dot{u}_n=i\,\left( u_{n+1}+u_{n-1}-2u_n+|u_{n}|^{2}\left(u_{n-1}+u_{n+1} \right)\right)+f[u_n ], \ \ |f[u_n ]|\ll 1,
\eeq
or, in matrix form
\beq
\dot{L}_n=A_{n+1}L_n -L_n A_n +F[u_n ], \ \ \ \ F[u_n ]=\begin{pmatrix}0 & f[u_n ]\\-\overline{f[u_n ]} & 0 \end{pmatrix},
\eeq
the main spectrum evolves in time generically in a non integrable fashion, and the theory of perturbations of integrable nonlinear evolution equations is the proper tool to have an analytic description of the effect of such a small perturbation on the dynamics under scrutiny.  

From the variation of the monodromy matrix $T$ in (\ref{def_monodromy}):
\beq\label{def_delta_L}
\ba{l}
\delta T(\lambda,t) =\sum\limits_{n=1}^N L_N\dots L_{n+1}\left(\delta L_n\right) L_{n-1}\dots L_1, \\
\delta L_n =\begin{pmatrix}0 & \delta u_n \\
-\overline{\delta u_n} & 0
\end{pmatrix},
\ea
\eeq
it follows that the trace of such a variation can be expressed in terms of the so-called transition matrix
\beq\label{def_trans_matrix}
\hat T(n,m,\lambda,t):=\Psi_{n}(t,\lambda)\Psi^{-1}_{m}(t,\lambda)=L_{n-1}L_{n-2}\dots L_{m} 
\eeq
as follows
\beq\label{general_variation}
\ba{l}
\tr\left(\delta T(\lambda,t) \right)=\sum\limits_{n=1}^N \tr\left(L_{n-1}\dots L_1 L_N\dots L_{n+1}\left(\delta L_n\right) \right)\\
=\sum\limits_{n=1}^N \tr\left(L_{N+n-1}\dots L_{n+1}\left(\delta L_n\right) \right)=\sum\limits_{n=1}^N \tr\left(\hat T(N+n,n+1,\lambda,t)\delta L_n\right) ,
\ea
\eeq
where we have used first the permutation properties of the trace and then the periodicity of $L_n$. 

Equation (\ref{general_variation}) is valid for any variation; in our case $\delta u_n(t)=\dot{u}_ndt$, then $\delta T=\dot T dt$, $\delta L_n =\dot{L}_n dt$, and equation (\ref{general_variation}) becomes
\beq\label{deriv_trace_T}
\ba{l}
\tr\left(\dot T(\lambda,t) \right)=\sum\limits_{n=1}^N \tr\left(\hat T(n+N,n+1,\lambda,t)\dot{L}_n\right) \\
=\sum\limits_{n=1}^N \tr\left(\hat T(n+N,n+1,\lambda,t)\left[A_{n+1}L_n -L_n A_n +F[u_n ] \right]\right) \\
=\sum\limits_{n=1}^N \tr\left(\hat T(n+N,n+1,\lambda,t)F[u_n ]\right)=\sum\limits_{n=1}^N \tr\left(\hat T(n+N,n,\lambda,t)L^{-1}_n F[u_n ]\right),                            
\ea
\eeq
where we have used first the fact that the AL vector field $A_{n+1}L_n -L_n A_n$ is not responsible for the time evolution of $\tr T$, and second the definitions (\ref{def_delta_L}) and (\ref{def_trans_matrix}) of $\delta L_n$ and of the transition matrix.

Recalling  the relation (\ref{tilde_T vs_T}) between the trace of T and that of $\tilde T$, we obtain the time derivative of $\tr \tilde T$ in terms of the perturbation:
\beq\label{deriv_trace_tilde_T}
(\tr\tilde T)_t=\left(\frac{\tr(T(\lambda,t))}{\sqrt{\det T(\lambda,t)}}\right)_t=\frac{\tr\left(\sum\limits_{n=1}^N\left[2\hat T(n+N,n,\lambda,t)-\tr T(\lambda,t) \right]L^{-1}_n F[u_n ] \right)}{2\sqrt{\det T(t)}}.
\eeq

Equation (\ref{deriv_trace_tilde_T}) describes in a rather implicit and nonlinear way the time evolution of the main spectrum, due to the nonintegrable perturbation $F[u_n ]$. A crucial simplification arises from the observation that, in the AW recurrence, during the linear stages of MI, characterized by the background solution, $\tr \tilde T$ is essentially constant (see Figures \ref{trace1},\ref{trace2} below). Therefore the variation takes place in the finite intervals in which the AWs appear, described to leading order by the AW solution, and since the Narita solution is exponentially localized in time, one can replace the integral over the time interval of appearance by an integral over the real time axis.

In conclusion the variation of $\tr \tilde T$ at each appearance of the AW is described by the following basic formula of the perturbation theory
\beq\label{basic}
\Delta\left(\tr\tilde T(\lambda) \right)=\int\limits_{-\infty}^{\infty}\frac{\tr\left(\sum\limits_{n=1}^N\left[2\hat T(n+N,n,\lambda,t)-\tr T(\lambda,t) \right]L^{-1}_n F[u_n(t) ] \right)}{2\sqrt{\det T(\lambda,t)}}dt,
\eeq
where all the quantities inside the integral (the transition and monodromy matrices, $L_n$ and $u_n$) correspond to the AW solution (\ref{Narita}). For instance (see the Appendix for details):
{\footnotesize
\begin{equation}\label{T_final}
\ba{l}
\hat T(n+N,n;\lambda_1)=\\
=\left(1+\,|a|^2\right)^{\frac{N}{2}}\left[-1+\frac{2N|a|\sqrt{1+|a|^2}\sin^2\phi\cos\frac{k}{2} \;g_n(t)}{\cos^2\phi\; H_n(t)}\begin{pmatrix}
\overline{q_1(n)}\overline{q_2(n)}&- \overline{q_2(n)}\overline{q_2(n)}\\[2mm]
\overline{q_1(n)}\overline{q_1(n)}&-\overline{q_1(n)}\overline{q_2(n)}
\end{pmatrix} \right]\;,
\ea
\end{equation}}
where
\beq\label{def_q1_q2}
\ba{l}
{q_1}_n=2\cos\left(\frac{\kappa (n-x_1)+i\eta\sigma (t-t_1) -\phi}{2} \right)e^{i\eta|a|^2 t+i\frac{ \arg a}{2}+i\nu t}, \\
{q_2}_n=-2\sqrt{\eta}\sin\left(\frac{\kappa (n-x_1)+i\eta\sigma (t-t_1) +\phi}{2} \right)e^{-i\eta|a|^2 t-i\frac{ \arg a}{2}+i\nu t},
\ea
\eeq
and
\begin{equation}\label{def_gn}
g_n(t)=q_1(n)q_1(n)e^{-2i|a|^2t}+q_2(n)q_2(n)e^{2i|a|^2t}+2 q_1(n)q_2(n)\sin\phi=4\cos^2\phi\;,
\end{equation}
{\scriptsize
\begin{equation}\label{def_Hn}
\ba{l}
H_n(t)=\left(\frac{|q_1(n)|^2}{\lambda_1}+\lambda_1|q_2(n)|^2\right)\left(\lambda_1|q_1(n)|^2+\frac{|q_2(n)|^2}{\lambda_1}\right)=\\
=16(1+|a|^2)\left[\cos\frac{k}{2}\cosh(\sigma (t-t_1))-\cos\left(k\left(n-x_1\right)\right)\sin\phi\right]\left[\cos\frac{k}{2}\cosh(\sigma (t-t_1))-\cos\left(k\left(n-x_1-1\right)\right)\sin\phi\right].
\ea
\end{equation}}
 
\begin{figure*}[h!]
\centering
\includegraphics[width=14cm,height=5.1cm]{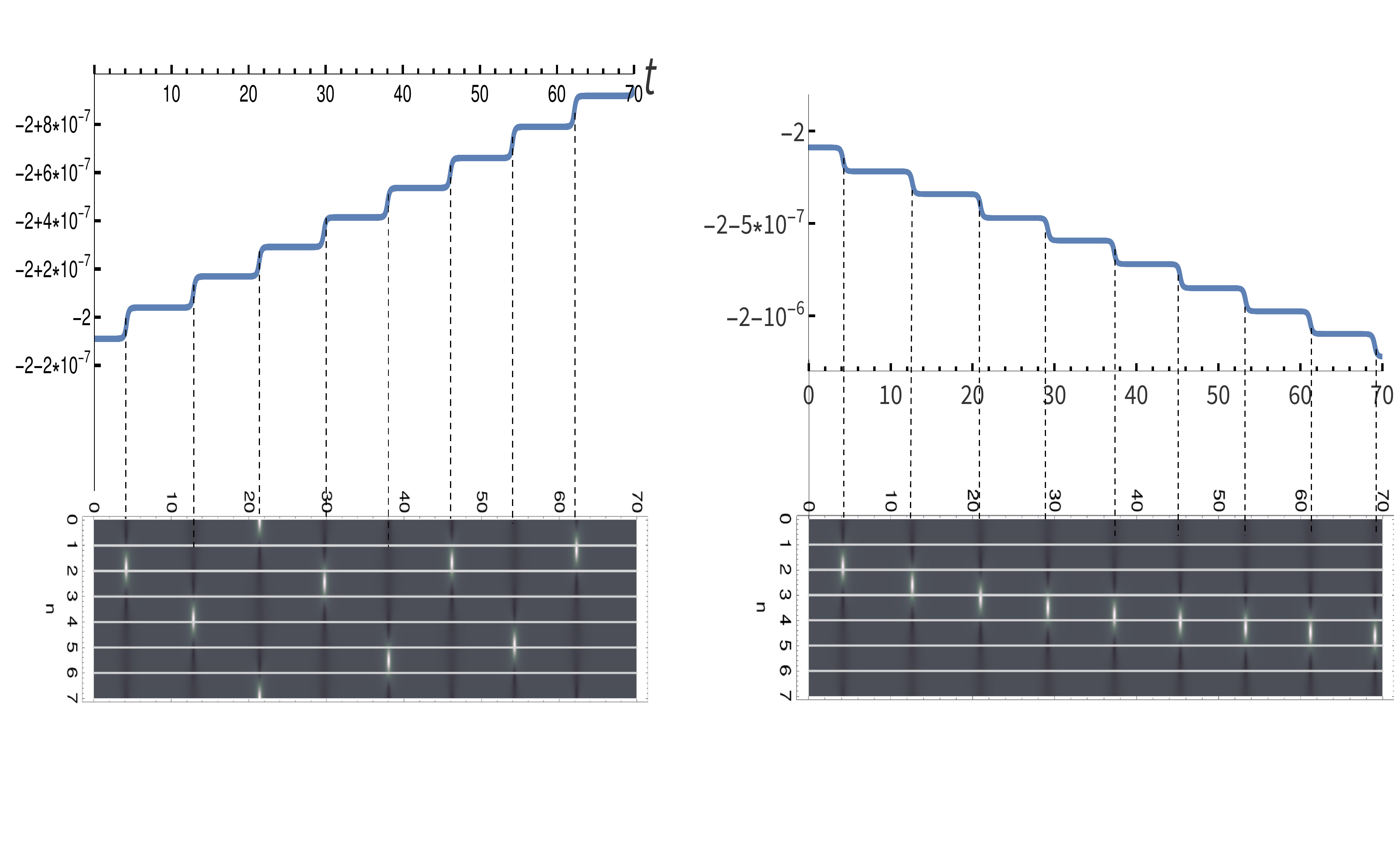}
\caption{\label{trace1} Dynamics of the real part $\Re\left(\frac{ Tr(T)}{\sqrt{\det T}}\right) $ of $Tr(\tilde T)$, in the presence of the small loss $\nu=10^{-8}$ or gain $\nu=-10^{-8}$. The variation corresponds to the AW appearance. The initial condition is: $u_n(0)=a(1+\epsilon(c_+ e^{ik_1 n}+c_- e^{-ik_1 n}))$, with $a=1.1$, $\epsilon=10^{-4}$, $k_1=\frac{2\pi}{N}$, $N=7$, $c_+=0.53-i~0.86$ and $c_-=-0.26+i~0.22$.} 
\end{figure*}

On the other hand, reasoning as in \cite{CGS}, for $\lambda\sim\lambda_1$:
\beq
\tr\tilde T(\lambda)\sim \frac{\tr \;T(\lambda_1)}{\sqrt{\det T}}+
\left(\frac{N\sin\phi_1}{\lambda_1 \cos\phi_1}\right)^2(\lambda-\lambda_1)^2.
\eeq\\
Evaluating this formula at $\lambda = E^+(\lambda_1)$ and recalling that $E^+(\lambda_1)-E^-(\lambda_1)= 2(E^+(\lambda_1)-\nobreak \lambda_1)$ and $\tr\tilde T(E^+(\lambda_1))=-2$, we have\\
\beq
\ba{l}
\tr \tilde T(\lambda_1)= \frac{Tr \;T(\lambda_1)}{\sqrt{\det T}}\sim -2-\left(\frac{N\sin\phi_1}{2\lambda_1\cos\phi_1}\right)^2 (E^+(\lambda_1)-E^-(\lambda_1))^2,
\ea
\eeq
implying that
\beq\label{basic2}
\Delta\left(\tr \tilde T(\lambda_1) \right)\sim -\left(\frac{N\sin\phi_1}{2\lambda_1\cos\phi_1}\right)^2 \Delta(E^+(\lambda_1)-E^-(\lambda_1))^2
\eeq

Equations \eqref{basic} and \eqref{basic2} complete the perturbation theory. In the following we apply it to three physically relevant perturbations, a linear loss, a linear gain, and a quintic Hamiltonian perturbation.

\subsection{Linear loss or gain perturbations}

If the perturbation is a linear loss or gain:
\beq
\ba{l}
f[u_n ]=\nu u_n, \ \ |\nu|\ll 1, \ \ \ \mbox{loss}: \ \nu<0; \ \ \ \mbox{gain}: \ \nu>0, \\
\Rightarrow \ \ \ F[u_n ]=\nu U_n(t), \ \ \ \ U_n:=\begin{pmatrix}0 & u_n \\
-\overline{u}_n & 0
\end{pmatrix}, 
\ea
\eeq
we can express the variation of $\tr\tilde T$ after the $j^{th}$ appearance of the AW, through the analytic formula:

\beq\label{basic_loss}
\ba{l}
\Delta_j(\tr\tilde T)=\Delta_j\left(\frac{ Tr(T)}{\sqrt{\det T}}\right)=\frac{2N\sqrt{1+|a|^2}\sin^2\phi\cos\frac{k}{2}}{\cos^2\phi}\int_{-\infty}^{\infty}\sum_{n=1}^{N}\frac{g_n(t)h_n(t)}{H_n(t)(1+|u_n(t)|^2)}d t\\
=-\nu\frac{2N|a|^2\sin^2\phi\cos\left(\frac{k}{2}\right)}{(1+|a|^2)} I(\phi,x_j),
\ea
\eeq
where $g_n(t),H_n(t)$ are defined in \eqref{def_gn},
\begin{equation*}
\begin{split}
&h_n(t)=\lambda_1\overline{u}_n\overline{q_2}^2+\frac{u_n}{\lambda_1}\overline{q_1}^2=\\
&=-4|a|\sqrt{1+|a|^2}\left(\cos\frac{k}{2}\cos(2\phi)+\cos\left(k\left(n-x_j\right)-i\sigma t\right)\sin\phi\right)\times\\
&\hspace{4cm}\times\frac{\left(\cos\frac{k}{2}\cosh(\sigma t)-\cos\left(k\left(n-x_j-1\right)\right)\sin\phi\right)}{\left(\cos\left(\frac{k}{2}\right)\cosh(\sigma (t-t_1))-\sin\phi\cos\left(k(n-x_j)\right)\right)},
\end{split}
\end{equation*}\\
and
{\footnotesize
\begin{equation*}
I(\phi,x_j)=\int_{-\infty}^{\infty}\left[\sum_{n=1}^{N}\frac{\cos\frac{\kappa}{2}\cos(2\phi)+\sin\phi\cos\left(\kappa_1\left(n-x_j\right)\right)\cosh\left(\sigma_1 t \right)}{\left[\cos\frac{k}{2}\cosh(\sigma t)-\cos\left(k\left(n-x_j+1\right)\right)\sin\phi\right]\left[\cos\frac{k}{2}\cosh(\sigma t)-\cos\left(k\left(n-x_j-1\right)\right)\sin\phi\right]}\right]d t\; .
\end{equation*}}
Note that $1+|u_n(t)|^2$ can be written as:
\begin{equation*}
\begin{split}
1+|u_n|^2=(1+|a|^2)&\left[\cos\left(\frac{k}{2}\right)\cosh(\sigma (t-t_1))-\sin\phi\cos\left(k(n-x_1+1)\right)\right]\times\\[2mm]
&\times\frac{\left[\cos\left(\frac{k}{2}\right)\cosh(\sigma (t-t_1))-\sin\phi\cos\left(k(n-x_1-1)\right)\right]}{\left[\cos\left(\frac{k}{2}\right)\cosh(\sigma (t-t_1))-\sin\phi\cos\left(k(n-x_1)\right)\right]^2}.
\end{split}
\end{equation*}\\

Comparing equations (\ref{basic_loss}) and (\ref{basic2}) we infer that the effect of the j-th  AW appearance is to modify the gap according to the formula:\\
\begin{equation*}
\Delta_j (E^+(\lambda_1)-E^-(\lambda_1))^2=\nu \frac{8|a|^2\lambda_1^2\cos^2\phi\cos\frac{\kappa_1}{2}}{N(1+|a|^2)}\;I(\phi_1,x^{(j)}),
\end{equation*}
that, combined with 
\beq
(E^+(\lambda_1)-E^-(\lambda_1))^2\Big |_{t=0}=\frac{\eps^2 |a|^2 \lambda^2_1 \alpha_1\beta_1}{(1+|a|^2)\sin^2\phi_1}
\eeq
provides the analytic formula for the position of the gap after the $n^{th}$ appearance in terms of the initial data:
\beq\label{gap_dynamics}
(E^+(\lambda_1)-E^-(\lambda_1))_n^2=\frac{\eps^2 |a|^2 \lambda^2_1 \alpha_1\beta_1}{(1+|a|^2)\sin^2\phi_1}+ \nu \frac{8|a|^2\lambda_1^2\cos^2\phi\cos\frac{\kappa_1}{2}}{N(1+|a|^2)}\sum\limits_{j=1}^n I(\phi,x^{(j)}).
\eeq

We can also define the useful sequence of complex numbers $\{Q_n\}$
\beq\label{def_Qn}
Q_n=\frac{(1+|a|^2)\sin^2\phi_1}{\eps^2 |a|^2\lambda_1^2}(E^+(\lambda_1)-E^-(\lambda_1))^2_n,\hspace{1cm}Q_0=\alpha_1\beta_1,
\eeq\\
where the subscript indicates the index of the AW appearence, obtaining
\beq\label{Qn}
Q_n=\alpha_1\beta_1 +\frac{\nu}{\eps^2}\frac{8\cos^2(\phi)\sin^2(\phi)\cos\frac{\kappa_1}{2}}{N}\sum_{j=1}^{n}I(\phi_1,x^{(j)}), \ \ n\ge 0.
\eeq

From \eqref{spectral_recurrence} and \eqref{def_Qn} we can conveniently express the $x$-shifts and recurrence times $\Delta_n t$ in terms of the sequence $\{Q_n \}$ of complex numbers:
\beq\label{recurrence_perturbed_AL}
\ba{l}
\Delta_n t=t^{(n+1)}-t^{(n)} =\frac{1}{\sigma_1}\log\left(\frac{\sigma_1^4}{4\eps^2 |a|^8|Q_n|\cos^2\frac{\kappa_1}{2}}\right), \\
\Delta_n x=x^{(n+1)}-x^{(n)} =\frac{\arg(Q_n)}{\kappa_1}.
\ea
\eeq

\begin{figure*}[h!!!!]
\centering
\includegraphics[trim=9cm 0cm -8cm 0 ,width=18cm,height=9.5cm]{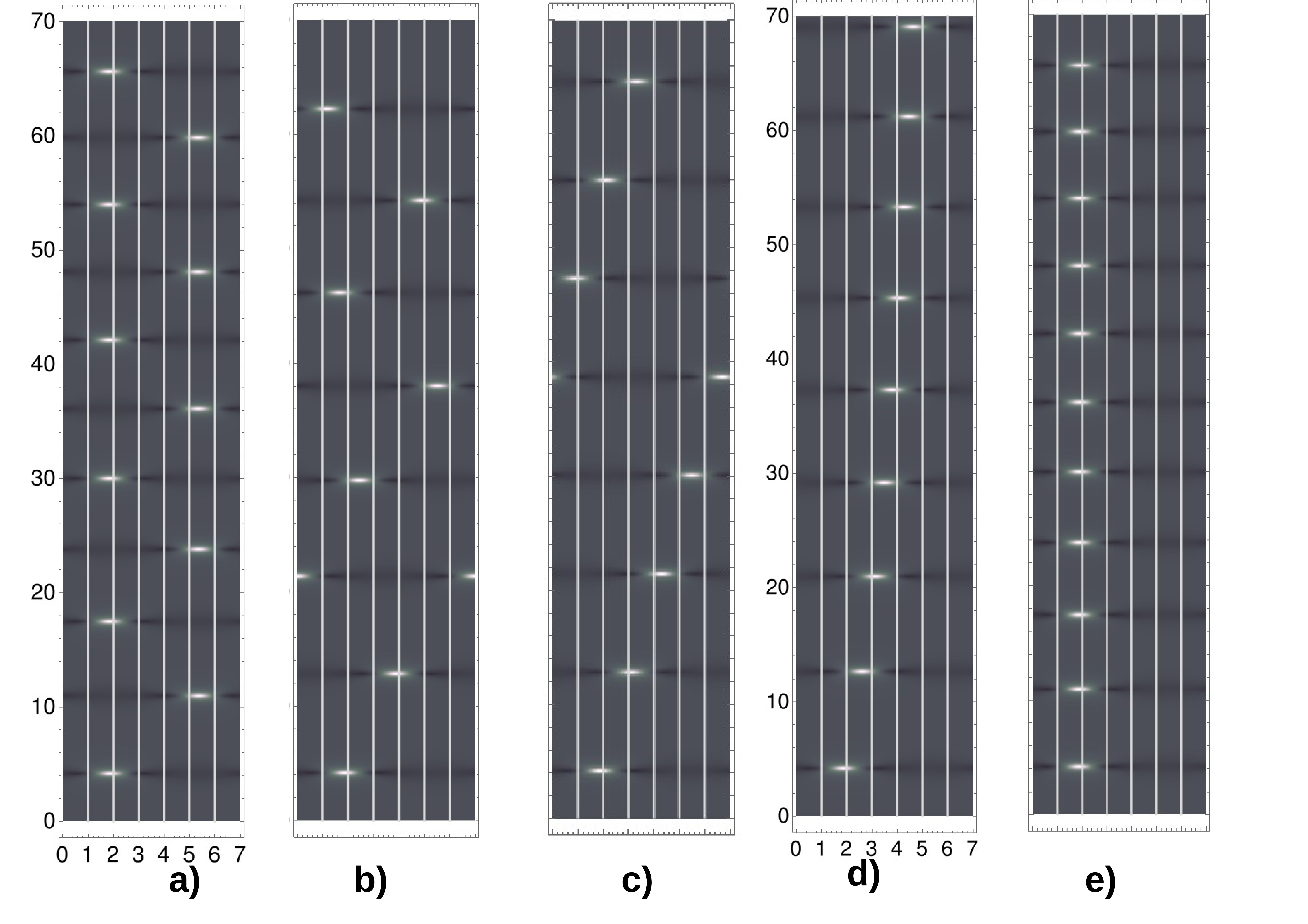}
\caption{\label{density_plots} Density plot of $|u_n(t)|$ with a small loss/gain. Initial condition $u_n(0)=a(1+\epsilon(c_+e^{ikn}+c_-e^{-ikn}))$, where $N=7$, $a=1.1$, $\epsilon=10^{-4}$, $c_+=0.53-i0.86$ and $c_-=-0.26+i0.22$ (as in Fig~\ref{fig:recurrence}). The integration is performed using the 6th order Runge-Kutta \cite{RungeKutta}.  \textbf{a)} $\nu=-10^{-6}$; the system enters immediately the attractor of the loss dynamics characterized by $\Delta x=N/2$.\textbf{b)} $\nu=-10^{-8}$; the system enters the loss attractor after a transient of few appearences. \textbf{c)} $\nu=0$; pure AL dynamics. \textbf{d)} $\nu=10^{-8}$; the system enters the attractor of the gain dynamics, characterized by $\Delta x=0$, after a transient of few appearences.  \textbf{e)} $\nu=10^{-6}$; the system enters immediately the attractor of the gain dynamics. } 
\end{figure*}

Summarizing the results of this section, the FPUT recurrence in the presence of a small loss or gain is described as follows. \\
\ \\
\textbf{Main result} {\it Consider the periodic Cauchy problem (\ref{Cauchy}) with period $N\in \NN^+$ for the AL equation perturbed by a small linear loss or gain
\beq\label{pert_AL}
i\, \dot{u}_n-2u_n+\left(1+\eta |u_{n}|^{2}\right)\left(u_{n-1}+u_{n+1} \right) =i\nu u_n, \ \ \nu\in\RR, \ |\nu|\ll 1,
\eeq
in the finite time interval $[0,T]$, in the simplest case of one unstable mode only ($\frac{2\pi}{\kappa_a} <N<\frac{4\pi}{\kappa_a}$). Then the solution is given, to leading order and up to $O(\eps)$ corrections, by the same analytic expression describing the AL FPUT recurrence (\ref{FPUT_AL}): 
\beq
u(x,t)=\sum\limits_{j=0}^n{\cal N}(x,t;\kappa_1,x^{(j)},t^{(j)},\rho_j,1)-ae^{2i|a|^2t}\frac{1-e^{4i\phi_1 n}}{1-e^{4i\phi_1}}+O(\eps),
\eeq
where the position and time of the first appearance $(x^{(1)},t^{(1)})$ of the AW, described by the Narita solution (\ref{Narita}), are essentially the same as in the unperturbed case and described by equations (\ref{def_recurrence}),(\ref{def_alpha_beta}), while the position and time of the subsequent appearances of the AWs, again described by the Narita solution (\ref{Narita}), are given by formulas (\ref{recurrence_perturbed_AL})} (see Figures~\ref{density_plots}).\\
\ \\

We first remark that, as for the NLS case, to obtain these results, we have implicitly assumed that
\beq\label{a_const}
|\nu|T, \ |a|^2 |\nu|T^2\ll 1 ,
\eeq
and the meaning of these conditions can be explained observing that the background solution
\beq\label{background_dissipation}
\tilde u_0(t,\nu)=a \exp(-\nu t)\exp\left(i\frac{|a|^2}{\nu}\left(1-\exp(-2\nu t) \right) \right)
\eeq
of (\ref{pert_AL}) behaves as follows
\beq
\ba{l}
\tilde u_0(t,\nu)=a\exp(\nu t)\exp\left(2i|a|^2t\left(1+\nu t+O(\nu t)^2\right) \right), \ \ |\nu|\ll 1 .
\ea
\eeq
Therefore the amplitude and the oscillation frequency of the background slowly decrease if $\nu<0$ (loss), and slowly increase if $\nu>0$ (gain). The condition $|\nu| T\ll 1$ means that we can neglect the slow decay/growth of the amplitudes of the background and of the AWs; the condition $|a|^2 |\nu| T^2 \ll 1$ means that we can neglect the slow decay/growth of the oscillation frequency and its effects. In particular,  $a$ can be treated as a constant parameter under the above assumptions.

We also remark that, \textit{unlike the unperturbed AL case, and unlike the perturbed NLS case, the $x$-shifts $\Delta_n x$ and the recurrence times $\Delta_n t$ after the $n^{th}$ appearance depend, through $Q_n$, on the positions $x^{(j)},~j=1,\dots,n$ of the first $n$ AW appearances}.\\
\ \\
In addition, since the functions $I(\phi_1,x^{(j)})$ are real and positive, the second term in the expression (\ref{Qn}) of $Q_n$ consists of a sum of positive terms in the case of a small gain ($\nu>0$), or of negative terms in the case of a small loss ($\nu<0$), and this second term becomes dominant in the sum:
\beq
Q_n \sim \frac{\nu}{\eps^2}\frac{8\cos^2(\phi)\sin^2(\phi)\cos\frac{\kappa_1}{2}}{N}\sum_{j=1}^{n}I(\phi_1,x^{(j)}), \ \ n\gg 1
\eeq
as $n$ increases. Therefore the AW recurrence tends to a lower dimensional asymptotic state (an attractor) characterized by the condition
\beq\label{attractor}
\ba{l}
\Delta x=0, \ \ \ \ \mbox{if} \ \ \ \ \nu>0, \ \mbox{small gain}, \\
\Delta x=\frac{N}{2}, \ \ \ \ \mbox{if} \ \ \ \ \nu<0, \ \mbox{small loss}.
\ea
\eeq

From (\ref{Qn}) we distinguish three different situations.\\
1) If $|\nu |\gg \eps^2$, after the first appearance, essentially the same as in the unperturbed case, the dynamics enters immediately one of the two attractors, depending on the sign of $\nu$ (see Figures~\ref{density_plots}~ \textbf{a)} and \textbf{e)}).\\
2) If $|\nu |=O(\eps^2)$, after the first appearance, essentially the same as in the unperturbed case, the dynamics reaches the above attractors after a suitable transient (see Figures ~\ref{density_plots}~ \textbf{b)} and \textbf{d)}). \\
3) If $|\nu |\ll \eps^2$, the dynamics is essentially the same as in the unperturbed case.\\
\newpage

To have an idea on how good this analytic theory is in describing the FPUT recurrence of AWs in the presence of loss or gain, in the following table we compare the numerical output of the experiment described in Figure~\ref{density_plots}~\textbf{b)} ($\eps=10^{-4},\nu=-10^{-8}$) with the above theoretical predictions:\\
\ \\
\begin{table}[h!]
\begin{center}
	%	\caption{More rows.}
	%	\label{tab:table1}
	\begin{tabular}{l l r}
		\hline
		\hline
		\hspace{5mm} 	 &    Numeric & Theor  \\
		\hline
		
		($\Delta X_1$, $\Delta T_1$) & (2.010914,$\;$ 8.671074 ) &                              (2.010913, $\;$8.671072)\\
		($\Delta X_2$, $\Delta T_2$) & (2.6311495,$\;$8.529947 ) &  
		(2.6311491,$\;$8.529944)\\
		($\Delta X_3$, $\Delta T_3$) & (2.9190295,$\;$8.380007) &
		(2.9190291,$\;$8.380003)\\
		($\Delta X_4$, $\Delta T_4$) & (3.0716356,$\;$8.256583 ) &  
		(3.0716351,$\;$8.256578)\\ 
		($\Delta X_5$, $\Delta T_5$) & (3.1626266,$\;$8.15582 ) &  
		(3.1626261,$\;$8.15681)\\ 
		\hline
	\end{tabular}
\end{center}
\end{table}\\
%\vspace{7cm} 

The whorst difference between numerics and the theory is in the $5^{th}$ decimal digit, while the expected error is $O(\eps)=O(10^{-4})$. Therefore this perturbation theory does even better than expected from theoretical arguments.

\begin{figure*}[h!!!!]
\centering
\includegraphics[trim=3.5cm 0cm -1cm 0 ,width=8cm,height=10cm]{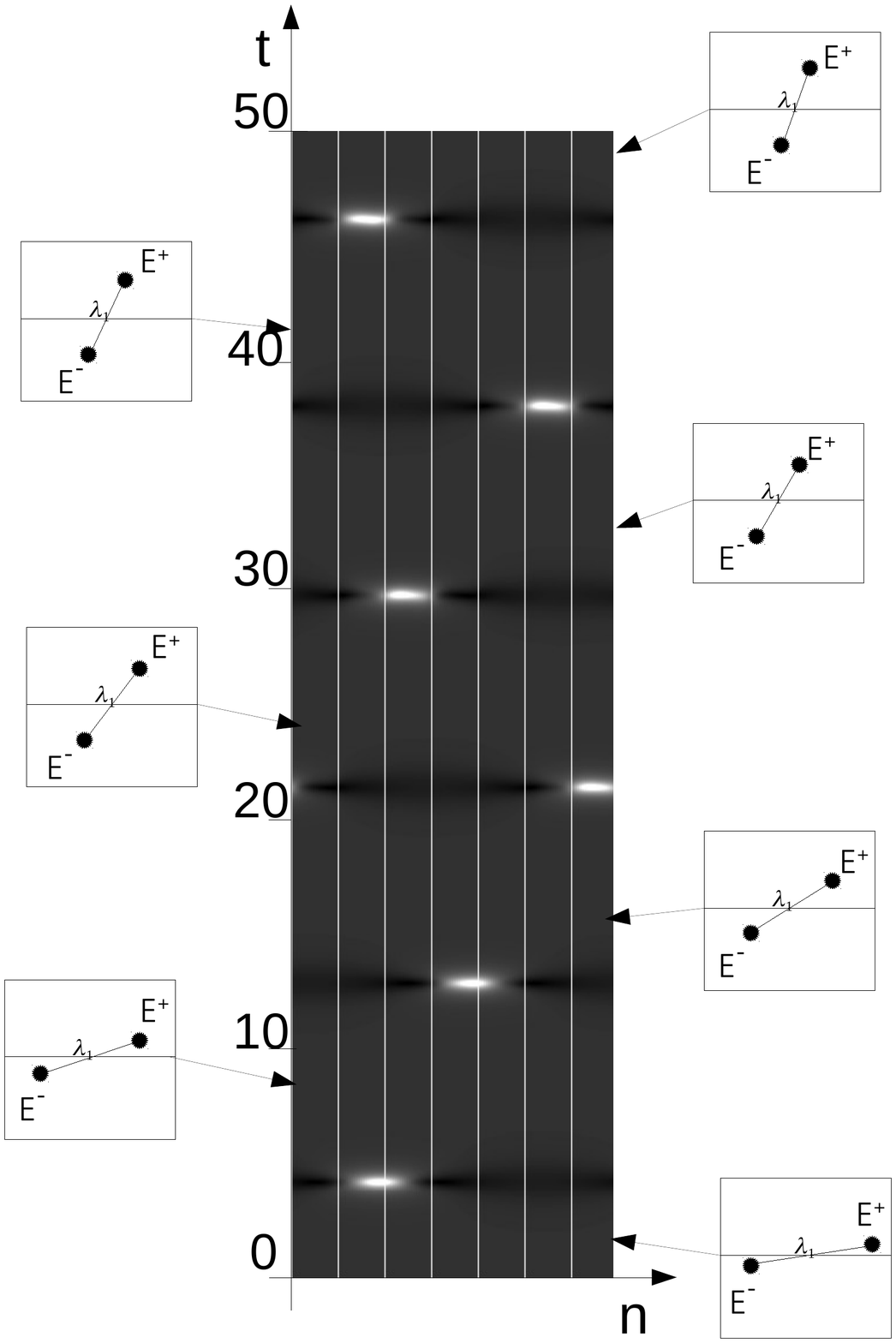}
\caption{Density plot of $|u_n(t)|$ and the (discrete) evolution of the gap $E^+-E^-$, due to the AW appearence, in the case of a small loss. The numerical experiment is the same as in Figure~\ref{density_plots}~\textbf{b)}.}\label{gaps_dyn} 
\end{figure*}

From (\ref{gap_dynamics}) it follows that
\beq
(E^+(\lambda_1)-E^-(\lambda_1))_n^2\to  \nu \frac{8|a|^2\lambda_1^2\cos^2\phi\cos\frac{\kappa_1}{2}}{N(1+|a|^2)}\sum\limits_{j=1}^n I(\phi,x^{(j)})
\eeq
as $n$ increases. Therefore the gap, whose initial inclination is arbitrary, tends to become horizontal in $\nu>0$ (gain), and vertical if $\nu<0$ (loss) (see Figure~\ref{gaps_dyn}).   
 \newpage
 \subsection{Hamiltonian perturbation}
 
In many physical contexts, a quintic term is introduced in NLS type models to describe higher order Hamiltonian effects. The matrix perturbation of the Lax pair is chosen as follows:\\ 
 \begin{equation}
 	F[u_n ]=-i\gamma\begin{pmatrix}0 & u_n \\\overline{u_n } & 0 \end{pmatrix}|u_n|^4, \ \ |\gamma|\ll 1,
      \end{equation}
and the variation of $\tr\tilde T$ after the $j^{th}$ appearance of the AW is described by the analytic formula
 {\small\begin{equation}\label{ham_int}
 		\begin{split}
 			&\Delta_j\left(\frac{ Tr(T)}{\sqrt{\det T}}\right)=-i\gamma\left(\frac{2N|a|^6\cos\phi_1\sin^2\phi_1\cos\frac{k_1}{2}}{1+|a|^2}\right)\tilde{I}(\phi_1,x^{(j)}),
 		\end{split}
              \end{equation}}
            
      {\small\begin{equation}\label{ham_int_def}
 		\begin{split}
 			& \tilde{I}(\phi_1,x^{(j)})=\int_{-\infty}^{\infty}\sum_{n=1}^{N}\frac{{\cal N}(n,t;x^{(j)})}{{\cal D}(n,t;x^{(j)})}d t,
 		\end{split}
 \end{equation}}
where
 {\scriptsize\begin{equation}
 		\begin{split}
 			&{\cal N}(n,t;x^{(j)})=\bigg|\cos\left(\frac{\kappa_1}{2}\right)\cosh(\sigma_1 (t)+2i\phi_1)+\sin\phi_1\cos\left(k_1 (n-x^{(j)})\right)\bigg|^4\times\\
 			&\times\bigg\{\sin(k_1(n-x^{(j)}))\left[2\cos\frac{k_1}{2}\sin^2\phi_1-\cos\frac{k_1}{2}\cosh(\sigma_1t)^2-\sin\phi_1\cos(k_1(n-x^{(j)}))\cosh(\sigma_1t\right]+\\
 			&+i\sinh(\sigma_1 t)\left[\cos\frac{k_1}{2}\cos(k_1(n-x^{(j)}))\cosh(\sigma_1 t-\sin\phi_1\bigg(\cos k_1+\sin^2(k_1(n-x^{(j)}))\bigg)\right]\bigg\},
 		\end{split}
 \end{equation}}
 and
 {\scriptsize\begin{equation}
 		\begin{split}
 			&{\cal D}(n,t;x^{(j)})=\left[\cos\frac{k_1}{2}\cosh(\sigma_1 t)-\cos\left(k_1\left(n-x^{(j)}\right)\right)\sin\phi_1\right]^5\times\\
 			&\times\left[\cos\frac{k_1}{2}\cosh(\sigma_1t)-\cos\left(k_1\left(n-x^{(j)}+1\right)\right)\sin\phi_1\right]\left[\cos\frac{k_1}{2}\cosh(\sigma_1t)-\cos\left(k_1\left(n-x^{(j)}-1\right)\right)\sin\phi_1\right].
 		\end{split}
 \end{equation}}

Proceeding as in the previous section, one obtains the same \textbf{main result} as before, but now

\beq\label{Qn_Ham}
Q_n=\alpha_1\beta_1 -i\frac{\gamma}{\eps^2}\frac{8 a^4\cos^3(\phi)\sin^2(\phi)\cos\frac{\kappa_1}{2}}{N}\sum_{j=1}^{n}\tilde{I}(\phi_1,x^{(j)}), \ \ n\ge 0.
\eeq

\begin{figure*}[h!!!!]
 	\centering
 	\includegraphics[trim=3cm 10cm 0cm 0 ,width=12cm,height=8cm]{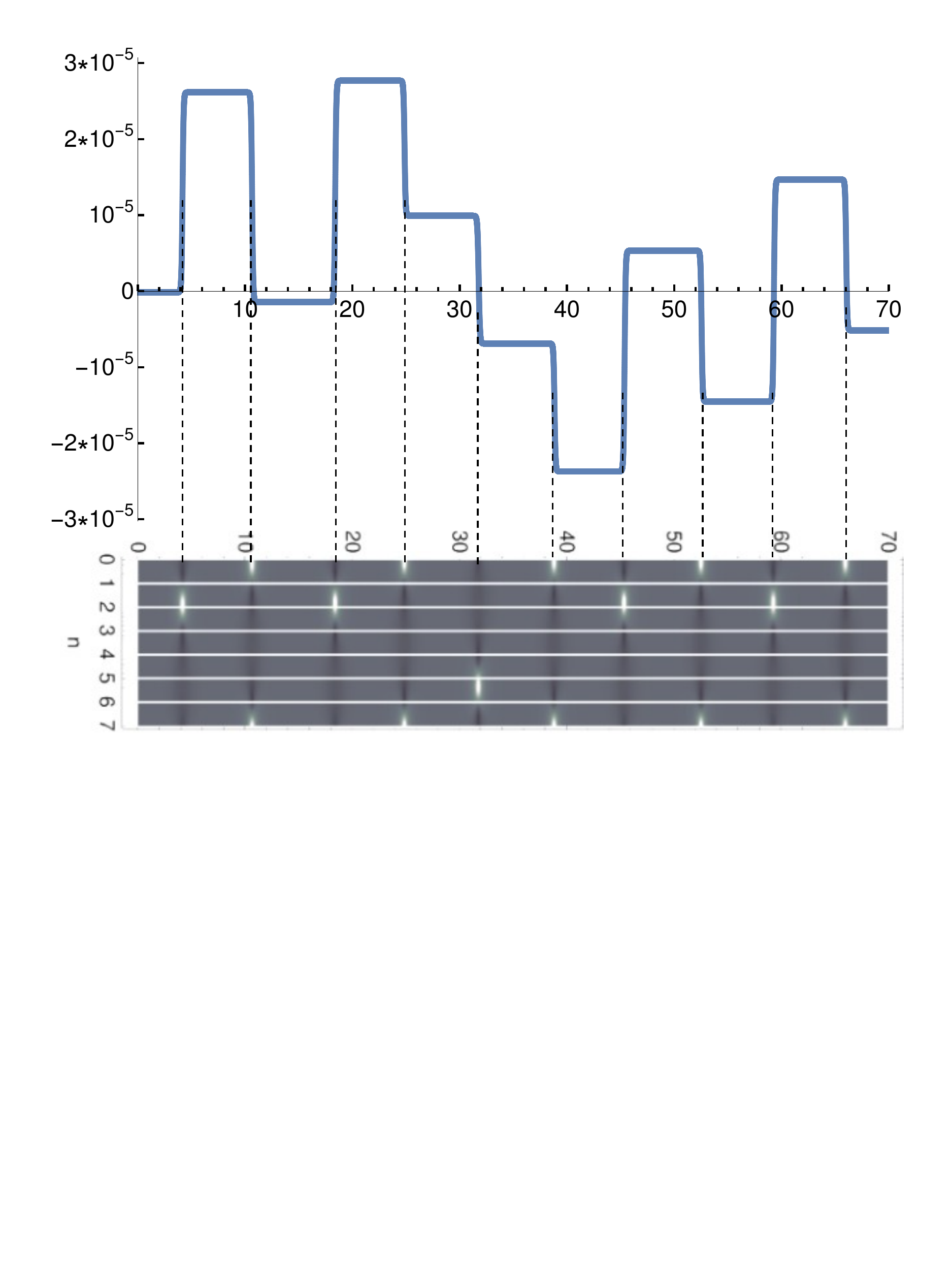}
 	\caption{Dynamics of the imaginary part of trace $\Im\left(\frac{ Tr(T)}{\sqrt{\det T}}\right) $, in the presence of the hamiltonian perturbation $\gamma=10^{-8}$. The variations correspond to the AW appearances. The initial condition is the same as before: $u_n(0)=a(1+\epsilon(c_+ e^{ik_1 n}+c_- e^{-ik_1 n}))$, with $a=1.1$, $\epsilon=10^{-4}$, $k_1=\frac{2\pi}{N}$, $N=7$, $c_+=0.53-i*0.86$ and $c_-=-0.26+i*0.22$.}\label{trace2} 
      \end{figure*}

As in the loss/gain case, the $x$-shifts $\Delta_n x$ and the recurrence times $\Delta_n t$ after the $n^{th}$ appearance depend, through $Q_n$ in \eqref{Qn_Ham}, on the positions $x^{(j)},~j=1,\dots,n$ of the first $n$ AW appearances. But now this contributions are purely imaginary and their imaginary parts can be either positive or negative. Therefore one cannot have, in general, asymptotic attractors. On the other hand, if $\gamma ~\eps^{-2}=O(1)$ as in Figure \ref{trace2}, then these purely imaginary terms prevail over the term $|\alpha_1\beta_1|=O(1)$ in \eqref{Qn_Ham}, due to the quintic perturbation. Therefore, at each appearance
\beq
\Delta_j x\sim \pm \frac{N}{4}
\eeq
(see Figure \ref{trace2}).

\newpage
      \iffalse
In the following table we compare quantitatively the above theoretical formulas with the numerical experiments. The worst difference is now in the $5^{th}$ decimal digit, while the estimated error would be in the $8^{}$ 
 \begin{table}[h!]
	\begin{center}
		%	\caption{More rows.}
		%	\label{tab:table1}
		\begin{tabular}{l l r}
			\hline
			\hline
			\hspace{5mm} 	 &    Numeric & Theor  \\
			\hline
			
			($\Delta X_1$, $\Delta T_1$) & (-1.746213,$\;$ 6.486309) &( 1.746245,$\;$6.486307)\\
			($\Delta X_2$, $\Delta T_2$) & ( 1.681436,$\;$ 7.743869) &( 1.681462,$\;$7.743848)\\
			($\Delta X_3$, $\Delta T_3$) & (-1.746426,$\;$ 6.461493) &(-1.746408,$\;$6.461495)\\
			($\Delta X_4$, $\Delta T_4$) & (-1.739963,$\;$ 6.908418) &(-1.739964,$\;$6.908414)\\ 
			($\Delta X_5$, $\Delta T_5$) & ( 1.735561,$\;$ 7.066487) &( 1.735566,$\;$7.066488)\\ 
        	($\Delta X_6$, $\Delta T_6$) & ( 1.745817,$\;$ 6.529162) &( 1.745803,$\;$6.529162)\\ 
            ($\Delta X_7$, $\Delta T_7$) & (-1.731339,$\;$ 7.178139) &(-1.731342,$\;$7.178136)\\ 
            ($\Delta X_8$, $\Delta T_8$) & ( 1.743162,$\;$ 6.741770) &( 1.743158,$\;$6.741766)\\  
            ($\Delta X_9$, $\Delta T_9$) & (-1.743229,$\;$ 6.737438)& (-1.743226,$\;$6.737435)\\         	
			\hline
		\end{tabular}
	\end{center}
\end{table}
\fi 
\section{Appendix}

To evaluate explicitly the right hand side of \eqref{basic}, the simplest way is to make use of the Darboux transformations (DTs) of the AL equations (see \cite{Geng}), concentrating on the $AL_+$ equation.\\
\ \\
\textbf{Proposition \cite{Geng}}. \textit{Let $\left(u_n^{[0]},{\underline\psi}_n^{[0]}(\lambda)\right)$ be a solution  of the Lax pair (\ref{AL_Lax_1}) for $\eta=1$, then $\left(u_n^{[1]},{\underline\psi}_n^{[1]}(\lambda)\right)$ is also a solution of (\ref{AL_Lax_1}), where
\beq\label{dressed}
\ba{l}
{\underline\psi}_n^{[1]}(\lambda)=D_n(\lambda){\underline\psi}_n^{[0]}(\lambda), \ \ \ u_n^{[1]}=a_{n+1}u_n^{[0]}+b_{n+1},
\ea
\eeq
\beq\label{def_D}
\ba{l}
D_n(\lambda)=\frac{1}{\lambda_1|q_2|^2+\frac{|q_1|^2}{\lambda_1}}\times\\
	\times\begin{pmatrix}
		\left(\frac{\lambda}{\lambda_1}-\frac{\lambda_1}{\lambda}\right)|{q_1}_n|^2+\left(\lambda\lambda_1-\frac{1}{\lambda\lambda_1} \right)|{q_2}_n|^2&-{q_1}_n{\overline{q}_2}_n\left(\lambda_1^2-\frac{1}{\lambda_1^2} \right)\\
		{q_2}_n{\overline{q}_1}_n\left(\lambda_1^2-\frac{1}{\lambda_1^2}\right)&-\left(\lambda\lambda_1-\frac{1}{\lambda\lambda_1} \right)|{q_1}_n|^2-\left(\frac{\lambda}{\lambda_1}-\frac{\lambda_1}{\lambda}\right)|{q_2}_n|^2
	\end{pmatrix},
      \ea
      \eeq
      \beq
      \ba{l}
      a_{n}=\frac{\lambda_1 |{q_1}_n|^2+|{q_1}_n|^2\lambda_1}{\lambda_1 |{q_2}_n|^2-{q_1}_n|^2/\lambda_1}, \ \ 
b_{n}=\frac{\lambda_1^2-1/\lambda_1^2}{\lambda_1 |{q_2}_n|^2-|{q_1}_n|^2/\lambda_1},     
\ea
\eeq
and ${\vec q}_n =({q_1}_n,{q_2}_n)^T$ is a linear combination of two independent solutions of (\ref{AL_Lax_1}) for $u_n=u_n^{[0]}$, evaluated at $\lambda=\lambda_1$, where $\lambda_1$ is a real parameter in the interval $(1,\lambda_0)$ (an unstable double point)} (see Figure \ref{spectrum1}).

Now we specialize this construction choosing $u^{[0]}_n(t)=a \,e^{2i\eta |a|^2 t}$ (the background solution); then a matrix fundamental solution of the Lax pair (\ref{AL_Lax_1}) reads
\beq\label{fund_2}
\ba{l}
{\Psi}^{[0]}_n=
\left(1+\eta\,|a|^2\right)^{\frac{n}{2}}e^{i\left(|a|^2t+\frac{\arg a}{2}\right)\sigma_3}\begin{pmatrix}
e^{i\frac{\eta}{2}(kn-\phi)-\frac{\sigma t}{2}}&&-e^{-i\frac{\eta}{2}(kn-\phi)+\frac{\sigma t}{2}}\\
\frac{i}{\sqrt{\eta}}e^{i\frac{\eta}{2}(kn+\phi)-\frac{\sigma t}{2}}&&\frac{i}{\sqrt{\eta}}e^{-i\frac{\eta}{2}(kn+\phi)+\frac{\sigma t}{2}}
\end{pmatrix}e^{i\nu\,t},
\ea
\eeq%\\[2mm]
where $\kappa=2\mu=\frac{2\pi}{N}$, and a linear combination of its columns leads to 
\beq\label{def_q1_q2}
\ba{l}
{q_1}_n=2\cos\left(\frac{\kappa (n-x_1)+i\eta\sigma (t-t_1) -\phi}{2} \right)e^{i\eta|a|^2 t+i\frac{ \arg a}{2}+i\nu t}, \\
{q_2}_n=-2\sqrt{\eta}\sin\left(\frac{\kappa (n-x_1)+i\eta\sigma (t-t_1) +\phi}{2} \right)e^{-i\eta|a|^2 t-i\frac{ \arg a}{2}+i\nu t},
\ea
\eeq
where $x_1,t_1$ are arbitrary real parameters.

Therefore \eqref{dressed} leads to the AW solution (\ref{Narita}):\\
\beq
u^{[1]}_{n}(t)=-a\;e^{2i\eta|a|^2t}\left[\frac{\cosh(\sigma (t-T)+2i\eta\phi)+\frac{\sin\phi}{\cos\left(\frac{\kappa}{2}\right)}\cos\left(\kappa (n-X)\right)}{\cosh(\sigma (t-T))-\frac{\sin\phi}{\cos\left(\frac{\kappa}{2}\right)}\cos\left(\kappa (n-X)\right)}\right],
\eeq
where $X=x_1-\frac{\eta}{2}-\frac{\pi}{2\kappa}$ and $T=t_1$. Correspondingly, the Darboux (dressing) matrix reads:
{\small
\beq\label{darb}
\ba{l}
D_n(\lambda)=\frac{1}{\lambda_1|q_2|^2+\frac{|q_1|^2}{\lambda_1}}\times\\
\times\begin{pmatrix}\left(\frac{\lambda}{\lambda_1}-\frac{\lambda_1}{\lambda}\right)|q_1|^2+\left(\lambda\lambda_1-\frac{1}{\lambda\lambda_1} \right)|q_2|^2&-q_1\overline{q}_2\left(\lambda_1^2-\frac{1}{\lambda_1^2} \right)\\
		q_2\overline{q}_1\left(\lambda_1^2-\frac{1}{\lambda_1^2}\right)&-\left(\lambda\lambda_1-\frac{1}{\lambda\lambda_1} \right)|q_1|^2-\left(\frac{\lambda}{\lambda_1}-\frac{\lambda_1}{\lambda}\right)|q_2|^2
\end{pmatrix}.
\ea
\eeq }

To calculate the remaining ingredients appearing in the basic formula (\ref{basic}), we observe that
\beq\label{transition}
\ba{l}
T^{[1]}(n+N,n;\lambda_1)=\Psi_{n+N}(\lambda_1)\Psi^{-1}_{n+N}(\lambda_1)=\displaystyle\lim_{\lambda\rightarrow\lambda_1}{D_{n+N}(\lambda)\,\Psi^{[0]}_{n+N}(\lambda_1)\left(\Psi_{n}^{[0]}(\lambda_1)\right)^{-1}D_{n}^{-1}(\lambda)}\\
=\displaystyle\lim_{\lambda\rightarrow\lambda_1}{D_{n+N}(\lambda)\,T^{[0]}(n+N,n,\lambda)D_{n}^{-1}(\lambda)}=\\
=\displaystyle\lim_{\lambda\rightarrow\lambda_1}\Big[\big(D_{n+N}(\lambda_1)+(\lambda-\lambda_1)\partial_\lambda D_{n+N}(\lambda)\big|_{\lambda_1}\big)\times\\
\times\big(T^{[0]}(n+N,n,\lambda_1)+(\lambda-\lambda_1)\partial_\lambda T^{[0]}(n+N,n,\lambda)\big|_{\lambda_1}\big)D_{n}^{-1}(\lambda)\Big].
\ea
\eeq
Therefore we need to evaluate the following quantities associated with the Darboux matrix \eqref{darb}:
\begin{equation}\label{darb_lambda1}
D_n(\lambda_1)=\frac{\left(\lambda_1^2-\frac{1}{\lambda_1^2}\right)}{\lambda_1|{q_2}_n|^2+\frac{|{q_1}_n|^2}{\lambda_1}}\begin{pmatrix}
\overline{{q_2}_n}\\\overline{{q_1}_n}
\end{pmatrix}(
{q_2}_n, -{q_1}_n) ,
\end{equation}
{\small
\begin{equation}\label{darb_der}
  \partial_\lambda D_n(\lambda)\bigg|_{\lambda=\lambda_1}=\frac{\mbox{diag}\left(2|{q_1}_n|^2+|{q_2}_n|^2\left(\lambda_1^2+\frac{1}{\lambda_1^2}\right),-2|{q_2}_n|^2-|{q_1}_n|^2\left(\lambda_1^2+\frac{1}{\lambda_1^2}\right) \right)}{\lambda_1\left(\lambda_1|{q_2}_n|^2+\frac{|{q_1}_n|^2}{\lambda_1}\right)},
\end{equation}}
{\small\begin{equation}\label{darb_inverse}
\ba{l}
D_n^{-1}(\lambda\sim\lambda_1)=\frac{\tilde{D}}{\lambda-\lambda_1}+O(1) , \ \
\tilde{D}:=\frac{\lambda_1}{2\left(\frac{|{q_2}_n|^2}{\lambda_1}+\lambda_1|{q_1}_n|^2\right)}\begin{pmatrix}
{q_1}_n\\{q_2}_n
\end{pmatrix}(
\overline{{q_1}_n}, -\overline{{q_2}_n}),
\ea
\end{equation} }
and those associated with the transition matrix of the background solution:
{\scriptsize
\begin{equation}\label{Tnm}
\ba{l}
T^{[0]}(n,m,\lambda,t)=\Psi^{[0]}_n(\lambda,t)\left(\Psi^{[0]}_m(\lambda,t)\right)^{-1}\\
=\left(1+\eta\,a^2\right)^{\frac{n-m}{2}}\begin{pmatrix}\cos(\mu (n-m)) +i\sin(\mu (n-m))\frac{ g_-+g_+}{ g_--g_+} && -2i\frac{\sin(\mu (n-m))}{ g_--g_+}e^{2i|a|^2t+i\arg a}\\[3mm]
2i\eta\frac{\sin(\mu (n-m))}{ g_--g_+}e^{-2i|a|^2t-i\arg a} && \cos(\mu (n-m)) -i\sin(\mu (n-m))\frac{ g_-+g_+}{ g_--g_+}
\end{pmatrix}\;,
\ea
\end{equation} }
{\scriptsize
\begin{equation}\label{T0_der}
\ba{l}
\partial_\lambda T^{[0]}(n,m;t)=\bigg[\frac{(n-m)\sin\phi}{\lambda\cos\phi}\begin{pmatrix}\sin(\mu (n-m)) -\cos(\mu (n-m))\frac{\sin\phi}{\cos\phi} & -\frac{\cos(\mu (n-m))}{\cos\phi}e^{2i|a|^2t}\\[3mm]
\frac{\cos(\mu (n-m))}{\cos\phi}e^{-2i|a|^2t} & \sin(\mu (n-m)) +\cos(\mu (n-m))\frac{\sin\phi}{\cos\phi}
\end{pmatrix}+\\[3mm]
+\frac{\sin(\mu (n-m))\cos\left(\frac{k}{2}\right)}{\lambda\sin\left(\frac{k}{2}\right)\cos^2\phi}\begin{pmatrix}1 & -\sin\phi\; e^{2i|a|^2t}+\\[3mm]
\sin\phi\;e^{-2i|a|^2t} &-1
\end{pmatrix}\bigg]\left(1+\eta\,a^2\right)^{\frac{n-m}{2}}\;.
\ea
\end{equation}}

At last, substituting these formulas in (\ref{transition}), we obtain $T^{[1]}(n+N,n;\lambda_1)$ as in \eqref{T_final}.
\vskip 10pt
\noindent
\textbf{Acknowledgments}\\
This research was supported by the Research Project of National Interest (PRIN) No. 2020X4T57A. It was also done within the activities of the INDAM-GNFM.

\end{document}